\documentclass[]{interact}

\usepackage{epstopdf}
\usepackage[caption=false]{subfig}

\usepackage[numbers,sort&compress]{natbib}
\bibpunct[, ]{[}{]}{,}{n}{,}{,}
\renewcommand\bibfont{\fontsize{10}{12}\selectfont}
\makeatletter
\def\NAT@def@citea{\def\@citea{\NAT@separator}}
\makeatother

\theoremstyle{plain}

\theoremstyle{definition}

\theoremstyle{remark}

\usepackage{tabularx} 
\usepackage{multirow}

\begin{document}

\title{Modelling the interaction between ethnicity and infectious disease transmission dynamics in Aotearoa New Zealand during the first Omicron wave of the COVID-19 pandemic}

\author{
\name{Vincent X. Lomas\textsuperscript{ab}\thanks{CONTACT Vincent X. Lomas Email: vincent.lomas@pg.canterbury.ac.nz}, and Tim Chambers\textsuperscript{b}, Michael J. Plank\textsuperscript{a}}
\affil{\textsuperscript{a}School of Mathematics and Statistics, University of Canterbury, Christchurch, Aotearoa; \textsuperscript{b}Ng\=ai Tahu Research Center, University of Canterbury, Christchurch, Aotearoa} 
}

\maketitle

\begin{abstract}
During the COVID-19 pandemic, Aotearoa followed an elimination strategy followed by a mitigation strategy, which saw high success and kept health impact low. However, there were inequities in health outcomes, notably that M\=aori and Pacific Peoples had lower vaccine coverage and experienced higher age-standardised rates of hospitalisation and death. Models provide predictions of disease spread and burden, which can effectively inform policy, but are often less good at including inequities/heterogeneity. Despite the differences in health outcomes, most models have not explicitly considered ethnic heterogeneities as factors. We developed such a model to investigate the first Omicron wave of the COVID-19 pandemic in Aotearoa, which was the first widespread community transmission of SARS-CoV-2. We analysed three models for contact patterns within and between ethnicities: proportionate, assortative, and unconstrained mixing, which were fit using ethnicity-specific data on reported cases and spatially disaggregated population counts. We found that M\=aori, Pacific, and Asian transmission rates were between 1.08-2.46, 1.50-3.89, and 0.80-0.92 times the European rates, respectively. We then found that from the parameters considered in the model, the disparity in ethnic transmission rates explained the majority of the observed ethnic disparity in attack rates, while assortativity and vaccination rates explained comparatively less.
\end{abstract} 

\section{Plain language summary}
Mathematical models of how an epidemic spreads through a population are important to inform policy decisions about public health interventions. Despite the relevance to disease spread, many models do not include important variables such as socioeconomic status and ethnicity. Here we investigate the first Omicron wave of COVID-19 in Aotearoa New Zealand with a model that splits the population into 10 groups relating to an individual’s disease and vaccination status: susceptible, susceptible, exposed, infectious, and recovered and unvaccinated, vaccinated, or vaccinated and booster. We further split the population into 4 ethnicity groups, corresponding to the main ethnicity classifications in Aotearoa: Māori, Pacific, Asian, and European/Other. We used this model to estimate how much contact rates would need to differ between ethnicities in order to explain observed differences in infection rates. We did this with three different assumptions about how many contacts occur with and between groups: (1) people mix with all ethnicities equally; (2) people are more likely to mix with others in the same ethnicity group; and (3) people are more likely to mix with others living in the same geographical area. From this, we estimated that Māori, Pacific, and Asian contact rates were 1.08-2.46, 1.50-3.89, and 0.80-0.92 times the European rates, respectively. We then found that the disparity in ethnic contact rates was sufficient to explain the majority of the observed disparity in infection rates by ethnicity, while preference for interacting with your own ethnic group and differences in vaccination rates between ethnicities explained comparatively less of the observed disparity.

\begin{abbreviations}
    SA1 - Statistical area 1, SA2 - Statistical area 2, CAR - Case Ascertainment Rate, PCR - Polymerase Chain Reaction, RAT - Rapid Antigen Test
\end{abbreviations}

\begin{keywords} 
mathematical modelling; epidemiology; health equity; COVID-19; compartmental model;
\end{keywords}

\section{Introduction}

When the COVID-19 pandemic began in 2020, it was predicted that M\=aori would be disproportionately affected \cite{jones_why_2020}, with one model estimating that the COVID-19 infection fatality rates would be 50\% higher for M\=aori compared to non-M\=aori \cite{steyn_estimated_2020}. A combination of living conditions, lack of access to quality health care, and existing health conditions led to M\=aori having a higher risk of infection and developing more serious symptoms of COVID-19 \cite{king_covid-19_2020}. During the first Omicron wave, M\=aori and Pacific Peoples had the highest per capita rates of cases and hospitalisations, with odds ratios of hospitalisation of 2.03 and 1.75 for M\=aori and Pacific Peoples, respectively \cite{whitehead_inequities_2023}. This was consistent with historical inequities in infectious disease burden experienced by Māori and Pacific Peoples \cite{rice_remembering_2019,wilson_differential_2012,esr_measles_2019,chapple_death_2018,bennett_rising_2021,cheung_severe_2024}. 

During the COVID-19 pandemic, Aotearoa changed its policy approach to COVID-19 at multiple stages. Aotearoa initially responded to the spread of COVID-19 with an elimination strategy. The elimination strategy aimed to completely eliminate COVID-19 from the country. This approach worked and resulted in about 30 cases and 0.4 deaths per hundred thousand population by August 2020 \cite{baker_successful_2020}. When vaccines became widely available, a mitigation strategy was used instead \cite{ministry_of_health_nz_managing_2024}. Starting in February, 2021, COVID-19 vaccines began to be administered to border and managed isolation and quarantine workers \cite{ardern_first_2021} and then the public thereafter. From March to May 2021, there was a large gap in the proportion of vaccinated M\=aori and Pacific Peoples compared to other ethnicities. At this time, the government implemented the mitigation strategy via the ``traffic light" system formally known as the COVID-19 Protection Framework. This framework implemented social distancing requirements, mask mandates, and COVID-19 vaccination certificates and their requirements \cite{parliamentary_counsel_office_covid-19_2021}. From May 2020 until the Omicron waves started in January 2022, there was very little community transmission. Starting in 2022, when there was high vaccination coverage \cite{ministry_of_health_new_2023}, there was some transmission of Omicron cases from isolation facilities at the border where international arrivals were being isolated. This resulted in a major wave that peaked on February 25th, 2022, with 24,226 daily confirmed cases. After this, the borders were reopened, and international travel was allowed in March 2022. Starting in July, an influx of immune-evasive variants was observed through the border. Most of these cases did not result in detected community transmission; a small amount did, resulting in large outbreaks \cite{douglas_tracing_2022}. Although the first and second doses of the vaccine had reduced efficacy against the Omicron variant compared to previous variants, it was shown that the booster doses were effective at stopping severe disease \cite{bowen_omicron_2022}.

Models add a lot of value to predictions of the spread and burden of disease and can effectively inform policy for prevention measures but are often not very good at including inequities/heterogeneity. Without these considerations, the models cannot address or predict these issues as effectively. Previous models have often struggled to account for ethnic heterogeneities despite their effect on disease spread \cite{williamson_opensafely_2020}, which has limited their ability to assess health equity. There are many models that use socioeconomic status \cite{goodfellow_covid-19_2024} or age \cite{lyra_covid-19_2020,miller_disease_2020,davies_age-dependent_2020} as factors, but models that consider ethnicity as a factor \cite{ma_modeling_2021,lee_ethnic_2023} are rarer. The mathematical models that were used to inform the Aotearoa policy approach to COVID-19 lacked the capability to adequately answer questions about health inequities and policy effectiveness. This limitation was identified as a major gap in the literature in these policy informing works \cite{datta_modelling_2023,rosenstrom_covsim_2024}. As such, there is a need to develop models that can answer these questions and consider a greater number of heterogeneities.

A method developed by Ma et al. considered ethnicity explicitly using data from Long Island and New York City during late April, 2020 \cite{ma_modeling_2021}. In this paper, they used a compartmental model and split the population into five ethnicity subgroups: non-Hispanic white, Hispanic or Latino, non-Hispanic Black, non-Hispanic Asian, and Multiracial/Other. Contact patterns were estimated using an ethnicity-specific contact rate, which was the average number of people interacted with each day, and an assortativity constant, $\epsilon$ , which measured the extent to which people are more likely to contact other people of the same ethnicity. Their model was first fit to data from a cross-sectional seroprevalence study from New York in late April 2020, using a proportionate mixing assumption ($\epsilon=0$), where people met according to population sizes and ethnicity-specific contact rates with no preference for interacting with any ethnicity. From this fit it was found that minorities had higher contact rates compared to non-Hispanic White people, with contact rates in New York City being 2.25, 1.62, 0.86, and 1.28 times the non-Hispanic White rate for the Hispanic or Latino, non-Hispanic Black American, non-Hispanic Asian, and Multiracial/Other groups, respectively.

Ma et al. then estimated a value for the assortativity constant based on the extent to which people of the same ethnicity tend to be clustered together in the same census blocks, with an assumption of density-based proportionate mixing within each block. This process gave more weight to the ethnic distribution of higher population blocks. The assortativity constant was chosen to minimise the difference between a census-informed matrix and a matrix formed using the total populations and the assortativity constant. The result of this fit was that 46\% and 39\% of an ethnicity's contacts were purely intra-ethnic contact in New York and Long Island, respectively; the disparity in ethnic contact rates was lowered to 1.62, 1.35, 0.90, and 1.17 times the non-Hispanic rate in New York City for the Hispanic or Latino, non-Hispanic Black, non-Hispanic Asian, and Multiracial/Other groups, respectively, but the disparity in rates was still present. 


Here, we applied a similar method to Ma et al. and aimed to estimate ethnicity-specific contact patterns by fitting a model to data on confirmed cases of COVID-19 during Aotearoa's first Omicron wave in 2022. We also aimed to try and understand the relative contributions of different drivers (vaccination rates, contact rates, and assortative contact patterns) to observed trends. We made two main adjustments to the Ma et al. method: we considered a vaccinated population in addition to an unvaccinated population and we changed the way contacts were estimated within census blocks, which allowed us to work directly with the census-informed matrix. We then applied our model to the first Omicron wave in Aotearoa to analyse differences in transmission rate by ethnicity. These transmission rates were then used as parameters to investigate various scenarios to analyse the spread of COVID-19 in the model.

\section{Methods}
In a closed, homogeneous population, an infectious disease can be modelled by a system of differential equations known as the SEIR model: 

\begin{align}
    \frac{dS}{dt} & = -\beta SI\\
    \frac{dE}{dt} & = \beta SI - \sigma E\\
    \frac{dI}{dt} & = \sigma E - \gamma I\\
    \frac{dR}{dt} & = \gamma I
\end{align}
where S, E, I, and R refer to the susceptible, exposed, infectious, and recovered populations, respectively. We set the mean latent period, $1/\sigma$, to 3 days and the mean infectious period, $1/\gamma$, to 4 days to match the Ma et al. study (A table of parameter values can be seen in Table \ref{tab:parameter_values}). There is some evidence that the Omicron variant had a shorter latent and infectious period than the wild-type variant \cite{park_inferring_2023}. However, as we are mainly interested in cumulative attack rates here, our results are not sensitive to choices for these parameters. The per capita transmission rate was defined in terms of the basic reproduction number, $R_0$, and the total population, $N$, $\beta=\gamma R_0/N$.

We then extended this model in the same way as Ma et al. to incorporate multiple ethnic groups by considering a per capita transmission matrix, $\boldsymbol{\beta}$, which governs the rate at which transmission happens within and between the considered ethnic groups. The per capita transmission matrix elements, $\beta_{ij}$, were the average proportion of susceptible people in ethnic group $i$ an infectious individual from ethnic group $j$ infected each day. In addition to this (beyond the Ma et al. extension), we also stratified the susceptible population according to whether they were: (i) unvaccinated ($S$); (ii) had received the primary vaccination course, normally two doses ($S_V$); (iii) had received a third vaccine dose, i.e. a booster ($S_{vb}$). There were relatively few people who received one dose and not a second, as such we grouped them into the primary vaccination  We created separate compartments to track when these populations become exposed and infectious ($\textbf{E}$, $\textbf{E}_v$, $\textbf{E}_{vb}$, $\textbf{I}$, $\textbf{I}_v$, and $\textbf{I}_{vb}$, respectively). We assumed that the effect of vaccinations was both a reduction in susceptibility and a reduction in transmission when infectious, and that the effect of vaccinations/boosters were constant over our study period. This leads to the following model:

\begin{align}
    \frac{d\textbf{S}}{dt} & = -\boldsymbol{\beta}(\textbf{I}+v_{t}\textbf{I}_v+v_{tb}\textbf{I}_{vb})\circ\textbf{S}\label{eq:S_vacc}\\
    \frac{d\textbf{S}_v}{dt} & = -v_a\boldsymbol{\beta}(\textbf{I}+v_{t}\textbf{I}_v+v_{tb}\textbf{I}_{vb})\circ\textbf{S}_{v}\\
    \frac{d\textbf{S}_{vb}}{dt} & = -v_{ab}\boldsymbol{\beta}(\textbf{I}+v_{t}\textbf{I}_v+v_{tb}\textbf{I}_{vb})\circ\textbf{S}_{vb}\\
    \frac{d\textbf{E}}{dt} & = \boldsymbol{\beta}(\textbf{I}+v_{t}\textbf{I}_v+v_{tb}\textbf{I}_{vb})\circ\textbf{S} - \sigma \textbf{E}\\
    \frac{d\textbf{E}_v}{dt} & = v_a\boldsymbol{\beta}(\textbf{I}+v_{t}\textbf{I}_v+v_{tb}\textbf{I}_{vb})\circ\textbf{S}_{v} - \sigma \textbf{E}_v\\
    \frac{d\textbf{E}_{vb}}{dt} & = v_{ab}\boldsymbol{\beta}(\textbf{I}+v_{t}\textbf{I}_v+v_{tb}\textbf{I}_{vb})\circ\textbf{S}_{vb} - \sigma \textbf{E}_{vb}\\
    \frac{d\textbf{I}}{dt} & = \sigma \textbf{E} - \gamma \textbf{I}\\
    \frac{d\textbf{I}_v}{dt} & = \sigma \textbf{E}_v - \gamma \textbf{I}_v\\
    \frac{d\textbf{I}_{vb}}{dt} & = \sigma \textbf{E}_{vb} - \gamma \textbf{I}_{vb}\\
    \frac{d\textbf{R}}{dt} & = \gamma (\textbf{I}+\textbf{I}_v+\textbf{I}_{vb})\label{eq:R_vacc}
\end{align}

In these equations, $1-v_a$ and $1-v_{ab}$ represent vaccine effectiveness against infection for those in the two-dose and three-dose compartments, respectively, while $1-v_t$ and $1-v_{tb}$ are the same for vaccine effectiveness against transmission in infected individuals; $\circ$ denotes element-wise multiplication; and \textbf{S}, $\textbf{S}_v$, $\textbf{S}_{vb}$, $\textbf{E}$, $\textbf{E}_v$, $\textbf{E}_{vb}$, \textbf{I}, $\textbf{I}_v$, $\textbf{I}_{vb}$, and \textbf{R} denote column vectors containing the compartmental populations for each group. The ethnic groups we considered were M\=aori, Pacific, Asian, and European/Other, as they are the major ethnic groups within Aotearoa and are consistently identified in health and population data and were identified in our data \cite{ministry_of_health_new_2023,statsnz_statistical_2020,te_whatu_ora_covid-19_2025,te_whatu_ora_populations_2023}.

Vaccines had reduced effectiveness against infection via Omicron and as such we needed to estimate their effectiveness \cite{cromer_predicting_2023}. Hao et al. \cite{hao_predicting_2025} estimated vaccine effectiveness as function of time since the second vaccine dose or the third vaccine dose. We did not model time-dependent immunity explicitly, but instead chose representative values corresponding to the timing of the vaccine rollout relative to the modelled wave. Since the primary vaccine rollout in Aotearoa occurred mainly in the last few months of 2021 \cite{ministry_of_health_new_2023}, we considered that those who had not received a booster doses would have about 120 days of immunity decay. The bulk of the booster rollout occurred in late 2021 to early 2022, just before the modelled wave, so we considered that those in the boosted compartment would have about 50 days of immunity decay. Values for vaccine effectiveness can be seen in Table \ref{tab:parameter_values}.

\subsection{Transmission matrix}

We defined the per capita contact matrix using a similar method to Ma et al. \cite{ma_modeling_2021}. We considered that on average, each ethnic group interacted with a different number of people per day; this was modelled with an ethnicity-specific contact rate, $\hat a_k$, where $k$ denoted the ethnicity. We assumed that the proportion $q$ of contacts resulting in infection was the same for all ethnicities. We define the transmission vector as $\textbf{a}=q\hat{\textbf{a}}$. Each component, $a_k$, of the transmission vector was the average number of people infected by an infectious individual of ethnicity $k$ in a fully susceptibility population. The differences in transmission rates between ethnicities are interpreted as being caused by differences in contact rates, i.e. social factors, not biological factors. Using this definition, $q$ and $\hat{a}_k$ were not separately identifiable, and so we expressed the model in terms of $\boldsymbol{a}$ and aimed to estimate $\boldsymbol{a}$ by fitting the model to data as described below.

\subsubsection{Proportionate mixing}
We first explored the simplest case that assumed each person's contacts were distributed across ethnicity groups in proportion to their population size. This defined each element of the per capita transmission matrix $\beta_{ij}$ as proportional to the transmission rates of two ethnicities multiplied together: 

\begin{align}
    \beta_{ij,\,\text{prop mix}} & = \frac{a_ia_j}{\sum_k a_k N_k} \label{eq:beta_prop_mix}
\end{align}
where $N_k$ is the population of ethnic group $k$. This left four parameters to estimate, the values of $a_k$.

\subsubsection{Assortative mixing}
We then considered a more realistic representation of transmission by accounting for preferential interaction within ethnicities with the introduction of the assortativity coefficient, $\epsilon \in [0,1]$, which was assumed to be the same for all ethnicities. When $\epsilon=0$ this corresponds to proportionate mixing, while $\epsilon=1$ is purely within group mixing, and intermediate values of $\epsilon$ interpolate between these two extremes. This helped to model community and family interactions that were primarily intra-ethnic interactions. We defined the assortative per capita transmission matrix as:

\begin{align}
    \beta_{ij,\text{ assort}} & = (1-\epsilon)\frac{a_ia_j}{\sum_ka_kN_k} + \epsilon \delta_{ij}\frac{a_i}{N_i} \label{eq:beta}
\end{align}

where $\delta_{ij}=1$ if $i=j$ and $\delta_{ij}=0$ otherwise. As can be seen, interchanging $i$ and $j$ did not alter the equation, which meant that the per capita transmission matrix was symmetric.

\subsubsection{Initialisation}
To initialise the model, we assumed that a fixed proportion $w$ of each ethnicity was exposed to the infection, with the remaining population in one of the susceptible compartments, giving the following initial conditions:
\begin{align*}
    S_{k}(0)&=(1-w)N_k - N_{vk}\\
    S_{vk}(0) & = N_{vk}-N_{vbk}\\
    S_{vbk}(0) & = N_{vbk}\\
    E_{k}(0)&=wN_k\\
    E_{vk}(0)&=E_{vbk}(0)=I_{k}(0)=I_{vk}(0)=I_{vbk}(0)=R_{k}(0)=0
\end{align*}
$N_{vk}$ refers to the vaccinated population of ethnic group $k$, and $w=0.0001$ was the initial proportion of the population exposed to the disease (giving approximately 500 initially exposed people).

\subsection{Parameter estimation and Data sources}

\subsubsection{COVID-19 case data}

\begin{figure}
    \centering
    \includegraphics[width=0.9\textwidth]{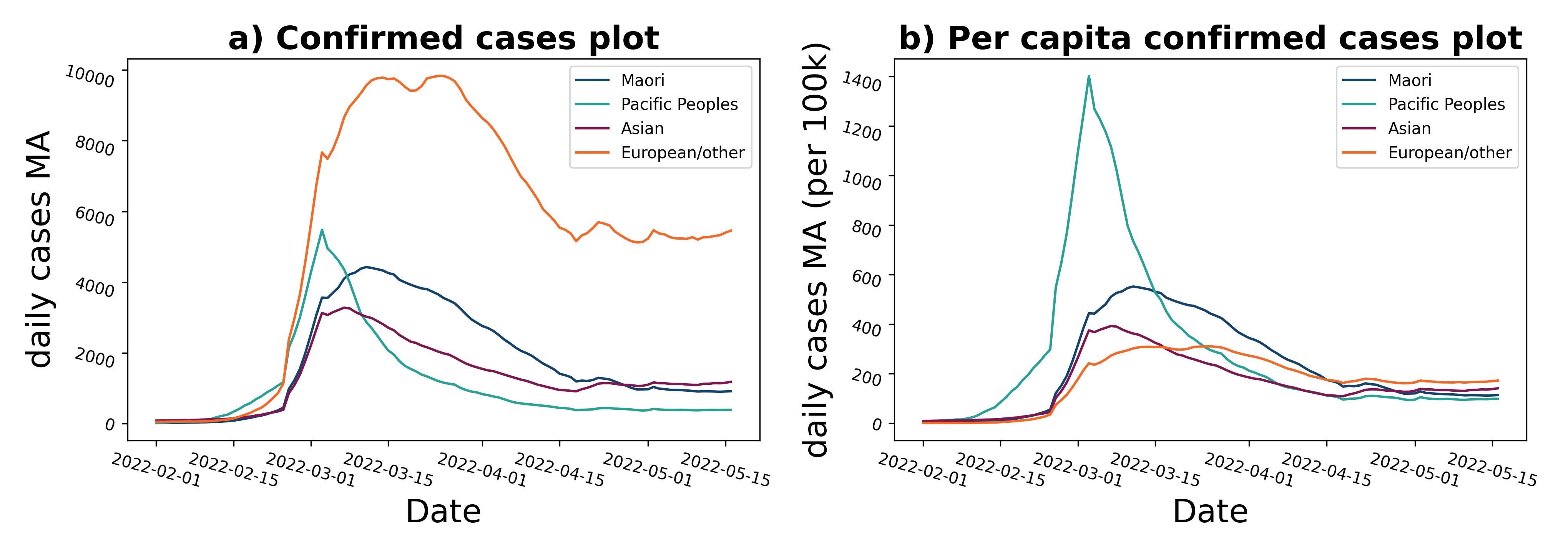}
    \caption{a) 7-day moving average of daily confirmed cases and b) 7-day moving average of confirmed cases per 100k for each ethnicity.}
    \label{fig:week_ma_raw}
\end{figure}

The data for the number of confirmed COVID-19 cases \cite{te_whatu_ora_covid-19_2025}. This data grouped the number of new cases per ethnicity for each day using a priority measure to assign each case to a single ethnicity (Figure \ref{fig:week_ma_raw}), which meant in the case of individuals belonging to more than one ethnic group, their case was assigned in the priority order of M\=aori, Pacific, Asian, and finally European/Other (values presented in Table \ref{tab:population_values}). This was opposed to a total measure that would count an individual's case towards all ethnic groups they identified with. The bulk of confirmed cases during the first Omicron wave were self-reported rapid antigen tests (RATs), with some polymerase chain reaction (PCR) tests (for cases presenting to healthcare). We assumed values for the case ascertainment rate (CAR), the proportion of cases that were reported, to account for asymptomatic cases and unreported cases (as the bulk of cases were self reported). As the true values for CAR were not known, we assumed four different situations: equal CARs of 40\%, 50\%, and 60\% across ethnicities, and a situation where the M\=aori and Pacific groups reported fewer cases and had a 40\% CAR, while the Asian and European/Other groups had a CAR of 60\%. We included this scenario as it is possible that Māori and Pacific Peoples had lower testing rates due to factors such as reduced access to healthcare services, higher rates of insecure employment, financial pressure, and/or inability to work remotely. CAR estimates were based on an Aotearoa study from February-July 2022 using wastewater surveillance. In this study, 40\% of 20–25-year-olds reported a case, giving an approximate lower bound on CAR, and model comparison to border workers gave an upper bound of around 60\% \cite{watson_jointly_2024}. It can be seen in Figure \ref{fig:week_ma_raw} that there was a sharp increase in the reported Pacific cases present when self-testing became available in late February 2022, which followed a period in which PCR testing capacity became overwhelmed \cite{allen__clarke_covid-19_2022}. This could imply that a significant portion of early Pacific cases could not be reported due to limited test availability and are therefore missing from the dataset.

\subsubsection{Vaccinated population data}
Ministry of Health data \cite{ministry_of_health_new_2023} was used to obtain values for the vaccinated populations for each ethnicity on the 23rd of February 2022 (see Table \ref{tab:population_values}). For simplicity, we placed people who had received one or two vaccine doses in the $S_v$ compartment, and those who had received a third dose in the $S_{vb}$ compartment. Note that the number of people who had received only one dose was relatively small (1.2\%) as most people completed the primary course of two doses. 

\subsubsection{Area-level population data}
Data for the population size of each ethnic group in specified geographical areas was needed to parameterise the assortative model. We used ethnicity-stratified population data at two levels of spatial resolution: Statistical Area 1 (SA1) and Statistical Area 2 (SA2) level. We assumed that the statistical area an individual lived in contained the bulk of interactions for that individual. StatsNZ reports population numbers at SA1 and SA2 level, which have typical populations of 100-200 and  1,000-4,000 people, respectively. The SA1 regions were formed by joining meshblocks (the smallest administrative unit) to ``allow the release of more detailed information about population characteristics..." \cite[][p. 13]{statsnz_statistical_2023}. The SA2 regions were defined in a way that ``aims to reflect communities that interact together socially and economically" \cite[][p. 14]{statsnz_statistical_2023}. As the SA2 blocks were formed to model a community's social interactions, they may appear a more accurate measure. However, a substantial amount of COVID-19 infections occurred within households \cite{chen_fangcang_2020}, and as such, the finer-scale grouping of the population gave more weight to local ethnic distributions and may give valuable insights. We used two datasets for SA2 population data with different population measures. The first was a dataset containing SA2 data by prioritised-ethnicity that assigned each individual to one ethnicity using the same priority as the COVID-19 case dataset. This data was downloaded from the Populations web tool (30 April, 2025), Health New Zealand, based on customised population projections provided by Stats NZ according to assumptions agreed to by Health New Zealand \cite{te_whatu_ora_populations_2023}; the other was a dataset from Stats NZ with a total measure available at both SA1- and SA2-level that counted an individual towards all ethnic groups that person identified with \cite{statsnz_statistical_2020}. Statistics about each dataset are present in Supplementary Table 1. Maps of the ethnic composition of each SA2 region are presented in Supplementary Figures 2 and 3. The Asian and Pacific populations are clustered primarily in Auckland, which is made up of small, densely population statistical areas, and as such the population appears much lesser than it is. These maps also show the high variability of the geographical areas of SA2s, which are mostly divided according to population.


The national ethnic population estimates (seen in Table \ref{tab:population_values}) were found by summing over the SA2 populations by prioritised ethnicity.

\begin{table}
\centering
\tbl{Table of parameter values.}{\begin{tabular}{lcr} 
\toprule
\textbf{Symbol} & \textbf{Meaning} & \textbf{Value}\\
\midrule
$\boldsymbol{\beta}$ & Per capita transmission matrix & Method dependent\\
\midrule
$\sigma$ & Inverse of latent period & 1/3 day$^{-1}$\\
\midrule
$\gamma$ & Inverse of infectious period & 1/4 day$^{-1}$\\
\midrule
$v_a$ & Complement of vaccine acquisition prevention effectiveness & 0.8\\
\midrule
$v_t$ & Complement of transmission prevention effectiveness & 0.95\\
\midrule
$v_ab$ & Complement of vaccine and booster acquisition prevention effectiveness & 0.55\\
\midrule
$v_tb$ & Complement of vaccine and booster transmission prevention effectiveness & 0.85\\
\bottomrule
\end{tabular}}
\label{tab:parameter_values}
\end{table}

\begin{table}
\centering
\centering
\tbl{Table of population characteristics}{\begin{tabular}{lccccc}
\toprule
\textbf{Symbol} & \textbf{Meaning} &  \multicolumn{4}{c}{\textbf{Value}}\\
& & \textbf{M\=aori} & \textbf{Pacific} & \textbf{Asian} & \textbf{European/Other}\\
\midrule
$N_k$ & Population of ethnic group $k$ & 888840 & 359480 & 820580 & 3048470\\
\midrule
$N_{vk}$ & Population of ethnic group $k$ with at least one vaccine (\%) & 58.0 & 76.5 & 76.8 & 84.0\\
\midrule
$N_{vbk}$ & Population of ethnic group $k$ with a booster (\%) & 20.9 & 30.9 & 40.1 & 51.7\\
\midrule
& cumulative reported cases per capita for ethnic group $k$ (\%)& 23.9  & 30.5 & 17.1 & 18.6\\
\bottomrule
\end{tabular}}
\label{tab:population_values}
\end{table}

\subsubsection{Proportionate mixing model estimation}
We first considered the case where people mixed with all ethnicities in proportion to their population size. Ethnicity-specific attack rates (i.e., cumulative number of cases per capita) were calculated from the 1st of February to the 16th of May, 2022, to capture the first Omicron wave in Aotearoa. These attack rates were then scaled up using the assumed CAR. The four ethnicity-specific transmission rates, $a_k$, were chosen such that the attack rates from the model matched the attack rates from the case data by using the optimize.least\_squares function from the SciPy package in Python with its default parameters. As there were 4 parameters and 4 attack rates to fit to, we found a unique solution for the transmission rates that provided an exact match to the target attack rates. All models were fit over a 365-day period, as it allowed sufficient time for the epidemic to run its course.

\subsubsection{Assortative mixing model estimation}
We then considered the case where people were more likely to mix with others in the same ethnic group. A method to determine the 5th parameter, the assortativity constant, was needed. Statistical area data was used to select a value for the assortativity constant using a similar method as Ma et al. \cite{ma_modeling_2021}. The method first constructed an empirically estimated contact matrix, $\boldsymbol{T}'$, using a specified transmission vector, $\boldsymbol{a}$, as defined in the following equation:

\begin{align}
    T_{ij}' & = \sum_l\frac{a_ia_jN_{i,l}N_{j,l}}{\sum_ka_kN_{k,l}} \label{eq:unconstrained_matrix}
\end{align}
where $N_{i,l}$ the population size of ethnicity group $i$ in statistical area $l$, and $T_{ij}'$ is the total number of contacts between people in group $i$ and people in group $j$ (multiplied by the transmission probability).

The definition of this matrix assumes that proportionate mixing occurs within each statistical area, and then aggregates the resulting contacts over the statistical areas. This was based on an approximating assumption that people mix proportionately with other people in the same geographical area and do not mix with people in other areas. This assumption was not entirely accurate to the real world but gave an approximate empirically based estimate of the average relative number of interactions an individual from a given ethnicity group would have.

When constructing this matrix, we assumed frequency-dependent interactions within statistical areas, whereas Ma et al. assumed density-dependent interactions in their block groups. Frequency-dependent interactions assume that the mean number of people an individual interacts with per day is independent of the population size in the census block or statistical area. In contrast, density-dependent interactions assume that the number of people an individual interacts with per day increases proportionally to the population size in the census block or statistical area in which they live. If the population blocks were all the same size, there would be no difference between these models, but when the blocks are different sizes, the density-based assumption would lead to an aggregated social contact matrix (after normalisation) that is disproportionately affected by the contact patterns within the larger blocks. The Aotearoa statistical areas have substantial variation in population size and, in this situation, we believe that the frequency-dependent paradigm is more realistic than the density-dependent one. This ignores systematic variations in contact rates with population density, for example between urban and rural areas, but these would also not be well captured by the density-dependent formulate of Ma et al. \cite{ma_modeling_2021} as the most density populated statistical areas are not necessarily the ones with the largest population size (see Supplementary Figures 2 and 3).

The values of $\boldsymbol{a}$ and $\epsilon$ were estimated jointly via the iterative method used by Me at al. \cite{ma_modeling_2021} as follows. For a given transmission vector, $\boldsymbol{a}$, we found the assortativity value, $\epsilon$, that minimised the difference between the empirical matrix $T'$ and the assortative mixing matrix $T$, defined by Equation \ref{eq:beta}.

We did this by solving the following optimisation problem for $\epsilon$:

\begin{align} 
\epsilon & = \mathop{\mathrm{argmin}}_{\hat\epsilon}\sum_{i,j}|T_{ij}-T'_{ij}|\\
    & =\mathop{\mathrm{argmin}}_{\hat\epsilon} \sum_{ij}\left|(1-\hat\epsilon)\frac{a_ia_jN_iN_j}{\sum_ka_kN_k}+\hat\epsilon\delta_{ij}a_iN_j -  \sum_l^L\frac{a_ia_jN_{i,l}N_{j,l}}{\sum_ka_kN_{k,l}}\right|\label{eq:epsilon_optimisation}
\end{align} 
where $N_{i,l}$ is the population of ethnic group $i$ living in geographic region $l$. Then, for this value of epsilon, we chose the values of $\boldsymbol{a}$ for which the model matched with the target attack rates in the same way as described previously. This iterative process was repeated until the absolute difference between successive values of $\epsilon$ was less than $10^{-6}$.

\subsubsection{Unconstrained mixing model estimation}
Finally, we considered a model in which people were more likely to mix with others living in the same geographical area, but without imposing an assortative mixing model, which constrains all ethnicity groups to have the same assortativity, and assigns between-group interactions in proportion to population size. We refer to this as the unconstrained mixing model. To implement this model, we directly defined the transmission matrix in terms of the area-level population data via Equation \ref{eq:unconstrained_matrix}. We then estimated the transmission rates as previously, by choosing the values of $a_i$ for which the model matched with the target attack rates.

Note that because the matrix in Equation \ref{eq:unconstrained_matrix} was formulated to have frequency-dependent contact rates, it has the required property that the the total average contact rate for an individual in group $i$ is equal to $a_i$, provided that $\sum_l N_{i,l}=N_i$ (see Supplementary Material section 3). This is not true of the density-dependent model used by Ma et al. \cite{ma_modeling_2021}, which can therefore only be used in conjunction with an assortative mixing assumption and not the unconstrained mixing model considered here. Because of the requirement that the area-level population data used to construct the matrix ($N_{i,l}$) aggregates to the national-level populations used in the model ($N_i$), we could only apply the unconstrained mixing model to the prioritised SA2 data and not the total ethnicity data.

\subsection{Scenario analysis}

We created four scenarios to quantify the effect that the ethnic disparity in vaccination rates, disparity in transmission rates, and the assortativity constant had on model results. These scenarios, seen in Table \ref{tab:quantisation}, took the assortative model fit and made one of the following changes: set $\epsilon=0$ to measure the effect of assortativity on the model; set all vaccination rates to the population-average vaccination rates; or set all transmission rates to the population-average transmission rates. These scenarios conserve the total number of vaccinations and the average transmission rate. 

Code and data to replicate results can be found in the GitHub repository \cite{lomas_modelling_2025}.

\begin{table}
\tbl{Table of scenarios to quantify the effects of parameters used in the model.}
{\begin{tabular}{cccc} 
\toprule
Scenario & Assortativity constant & Vaccination rate & Transmission rate\\
\midrule
1 & Fitted value & Ethnicity-specific & Ethnicity-specific\\
2 & Zero & Ethnicity-specific & Ethnicity-specific\\
3 & Fitted value & Population average & Ethnicity-specific\\
4 & Fitted value & Ethnicity-specific & Population average\\\bottomrule
\end{tabular}}
\label{tab:quantisation}
\end{table}

\section{Results}

\subsection{Proportionate mixing model}
\begin{figure}
\centering
\includegraphics[width=0.9\textwidth]{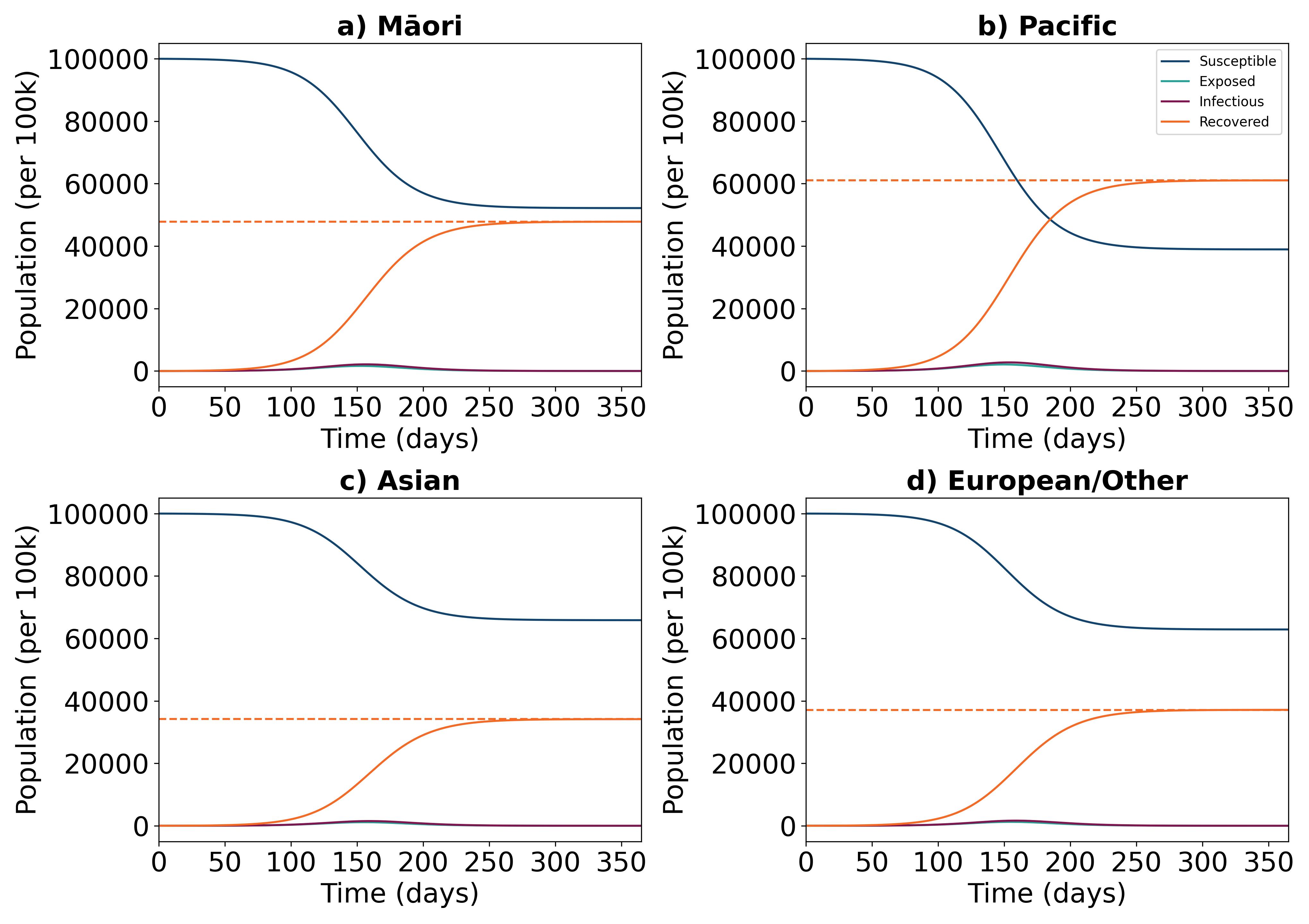}
\caption{Per capita SEIR plots fit to data on confirmed cases under the assumption of proportionate mixing and an assumed CAR of 50\% for a) M\=aori, b) Pacific, c) Asian, and d) Europeans/Other. The dashed line is the cumulative attack rate according to the data on reported cases and the assumed CAR of 50\%.}
\label{fig:SEIR_prop_mix}
\end{figure}

As expected with an SEIR model, a single epidemic wave that ends as a result of herd immunity in the population was observed (Figure \ref{fig:SEIR_prop_mix}). Recall that the transmission rates were chosen to exactly match the attack rates observed in the data (after adjustment for the assumed CAR). As such, the final recovered population in each ethnicity matched the number of people infected according to the attack rate data and the assumed value for CAR (indicated by the dashed horizontal line).

For all the assumed values of CAR, we chose transmission rates to exactly match the inferred attack rate values; see Figure \ref{fig:SA2_priority_assortative_unconstrained_transmission_rates} and Table \ref{tab:transmission_rates}. For all values of CAR investigated, the Pacific transmission rate was highest, followed by the M\=aori, European/Other, and then Asian rates, respectively. The initial reproduction number increased with a decrease in the assumed CAR and was highest when M\=aori and Pacific Peoples were given a higher CAR than the other groups.

\subsection{Assortative mixing model}
We now investigate the inclusion of assortativity in the model to represent preferential inter-ethnicity interactions. There was no immediately obvious choice for the value of assortativity, so we first explored model outputs over the full range of $0\leq \epsilon \leq 1$ (plots of this are presented in Supplementary Figure 1). We then estimated the value for assortativity and the transmission rates using the statistical area data (Figure \ref{fig:SA2_priority_assortative_unconstrained_transmission_rates}). We present the results for the SA2-prioritised data here; results for SA1/SA2 total ethnicity data are presented in the Supplementary Figure 5. The transmission rates estimated using the SA1-total and SA2-total data were qualitatively similar, but the SA1-total data had the highest value for $\epsilon$. This was an expected result due to smaller subdivisions sampling a smaller number of people, which increases the assortativity we would expect.

\subsection{Unconstrained mixing model}

\begin{figure}
\centering
\includegraphics[width=0.9\textwidth]{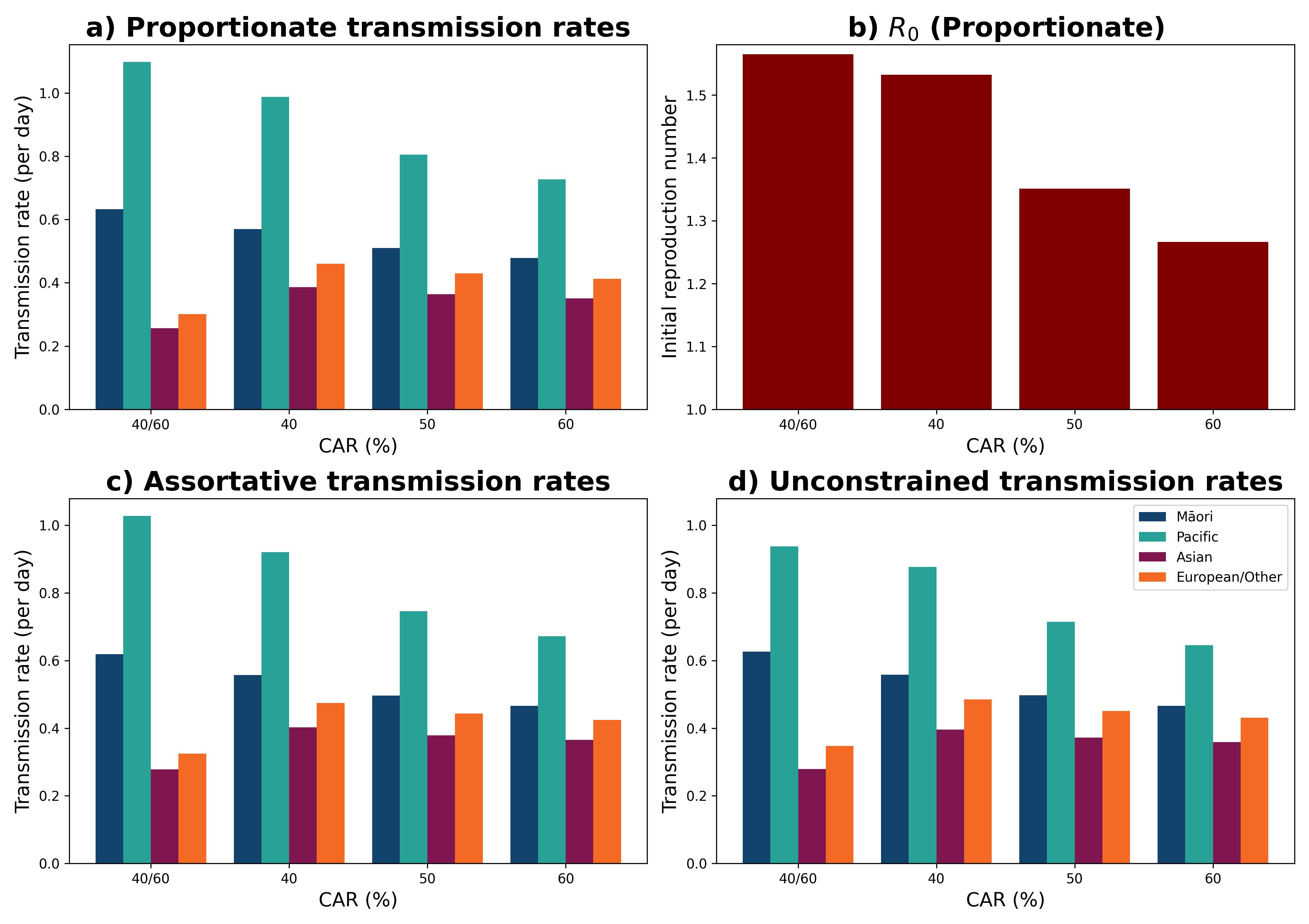}
\caption{
a) Transmission rates fit to the proportionate mixing model, b) the reproductive number of the proportionate mixing model, c) transmission rates fit to the assortative mixing model and d) transmission rates fit to the unconstrained mixing model. The transmission rates were chosen such that the attack rates from the model exactly matched the attack rates from the data, after adjustment with the assumed CARs. SA2 prioritised data was used for fitting the assortative mixing and unconstrained mixing models.}
\label{fig:SA2_priority_assortative_unconstrained_transmission_rates}
\end{figure}

We now construct the unconstrained mixing transmission matrix directly from the population data, without imposing a particular mixing pattern to eliminate assumptions about national proportional mixing and assortativity. Again, the transmission rates were chosen such that the attack rates from the model exactly match the data. The transmission rates can be seen in Figure \ref{fig:SA2_priority_assortative_unconstrained_transmission_rates}. When compared to the proportionate mixing rates, the unconstrained mixing model rates for M\=aori and Pacific rates were lower, while the Asian and European/Other rates were higher, which meant there was a lesser disparity between the ethnic transmission rates.

\begin{figure}
\centering
\includegraphics[width=\textwidth]{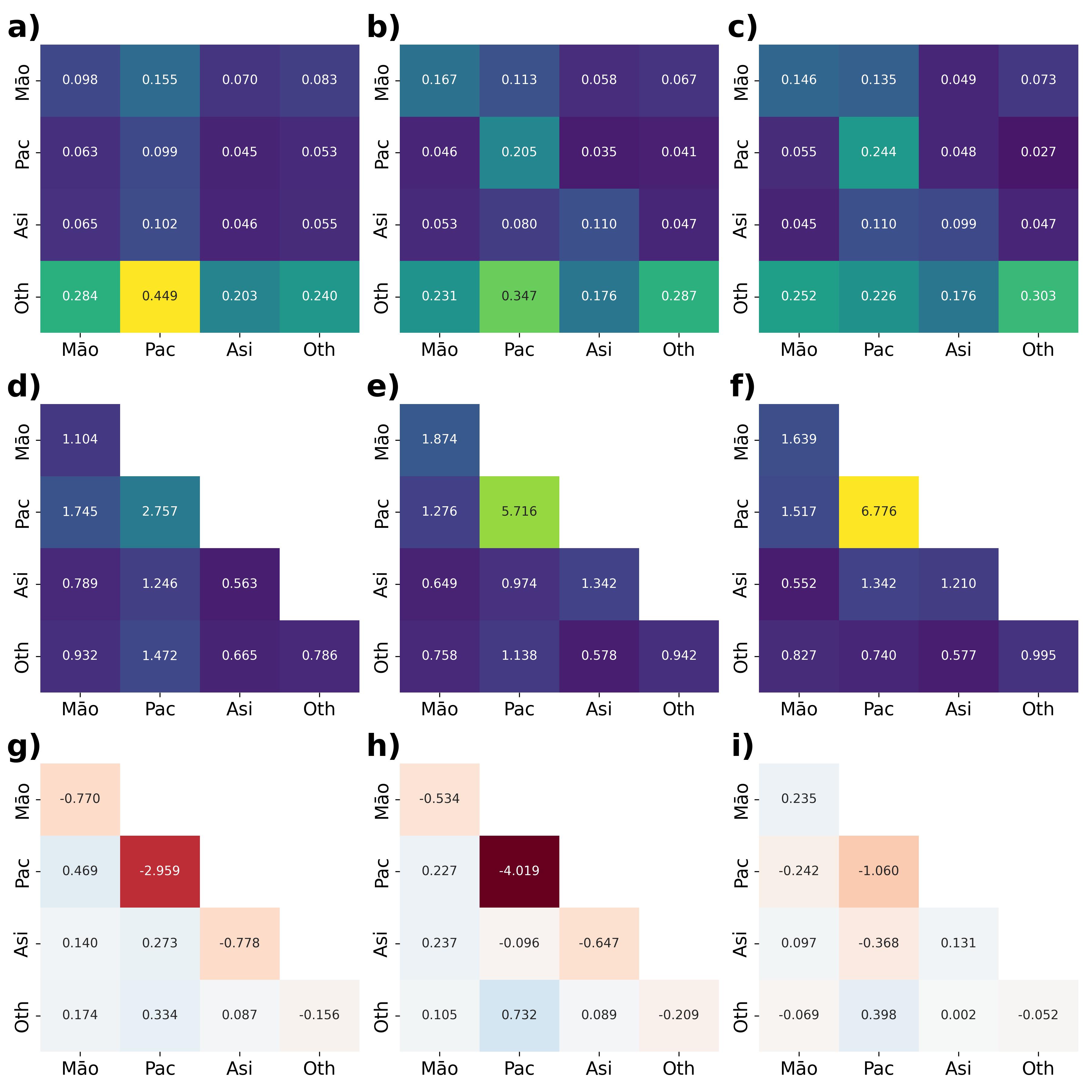}
\caption{A comparison between a) the proportionate mixing model transmission matrix; b) the assortative mixing model transmission matrix ($\epsilon=0.179$); c) the unconstrained mixing model transmission matrix; d) the proportionate mixing model per capita transmission matrix; e) the assortative mixing model per capita transmission matrix; and f) the unconstrained mixing model per capita transmission matrix; g) the difference between the proportionate and assortative mixing model per capita matrix; h) the difference between the proportionate and unconstrained mixing model per capita matrix; i) the difference between the assortative and unconstrained mixing model per capita matrix. Per capita matrix values are all scaled by $10^{-7}$}. Matrices were fit with the prioritised SA2 population data (where relevant) such that the attack rates from the model exactly matched the attack rates from the case data with an assumed CAR of 50\%. \label{fig:heatplot_transmission_comp_assortative_unconstrained}
\end{figure}

We compared the transmission matrices for the assortative and the unconstrained mixing models in Figure \ref{fig:heatplot_transmission_comp_assortative_unconstrained}. The matrices were relatively similar; the elements in the unconstrained mixing model matrix were 0.65-1.38 times the values in the assortative matrix, with the biggest difference between models being the frequency of interaction between the Pacific and the Asian and European/Other groups. All Pacific interactions were higher under the unconstrained model than the assortative model except for the Pacific-European/Other interaction, which was substantially lower. The diagonal elements differed less between the two models than the off-diagonal elements, with the unconstrained model diagonal elements being between 0.87 and 1.19 times the assortative matrix elements (with intra-Pacific interactions relatively increasing the most and intra-M\=aori interactions relatively decreasing the most). The transmission rates and initial reproduction numbers were similar for both models.

\begin{table} 
\centering
\tbl{Table of transmission rate values for prioritised SA2.}{\begin{tabular}{lc c c c c} 
\toprule
 &  & $a_k$ & & &\\ \cmidrule{3-6}
\textbf{CAR} & $\epsilon$ & M\=aori & Pacific & Asian & European/Other\\
\midrule
Proportionate Mixing &&&&&\\
\midrule
40/60 & & 0.633 & 1.098 & 0.257 & 0.302\\
40 & & 0.569 & 0.988 & 0.386 & 0.461\\
50 & & 0.510 & 0.805 & 0.364 & 0.430\\
60 & & 0.478 & 0.727 & 0.351 & 0.412\\
\midrule
Assortative Mixing\\
\midrule
40/60 & 0.145 & 0.619 & 1.027 & 0.279 & 0.325\\
40 & 0.176 & 0.557 & 0.920 & 0.402 & 0.475\\
50 & 0.183 & 0.497 & 0.746 & 0.379 & 0.443\\
60 & 0.186 & 0.466 & 0.672 & 0.365 & 0.424\\
\midrule
Unconstrained mixing\\
\midrule
40/60 & & 0.626 & 0.938 & 0.279 & 0.347\\
40 & & 0.558 & 0.877 & 0.396 & 0.485\\
50 & & 0.498 & 0.714 & 0.372 & 0.451\\
60 & & 0.466 & 0.645 & 0.359 & 0.431\\
\bottomrule
\end{tabular}}
\label{tab:transmission_rates}
\end{table}

\subsection{Sensitivity analysis}

To analyse the sensitivity of our model, we varied the assumptions and data sources we used (present in Supplementary Table 2). The choice of CAR, data source for spatially disaggregated population sizes by ethnicity, and mixing model did not change the qualitative results from fitting the model. The value of CAR determines the magnitude of ethnic disparity but does not change the fact that disparity is present. When we assumed a lower value for CAR rate, it lead to a greater disparity in the ethnicity-specific transmission rates, and when M\=aori and Pacific Peoples had a lower assumed CAR, the disparity in transmission rates was largest. Changing the assumed CAR did not change the relative ordering of transmission rates, i.e., Pacific Peoples always had the highest transmission rate. These results were also not highly sensitive to whether we used SA1 or SA2 data.

\subsubsection{Scenario analysis}

\begin{figure}
    \centering
    \includegraphics[width=0.7\linewidth]{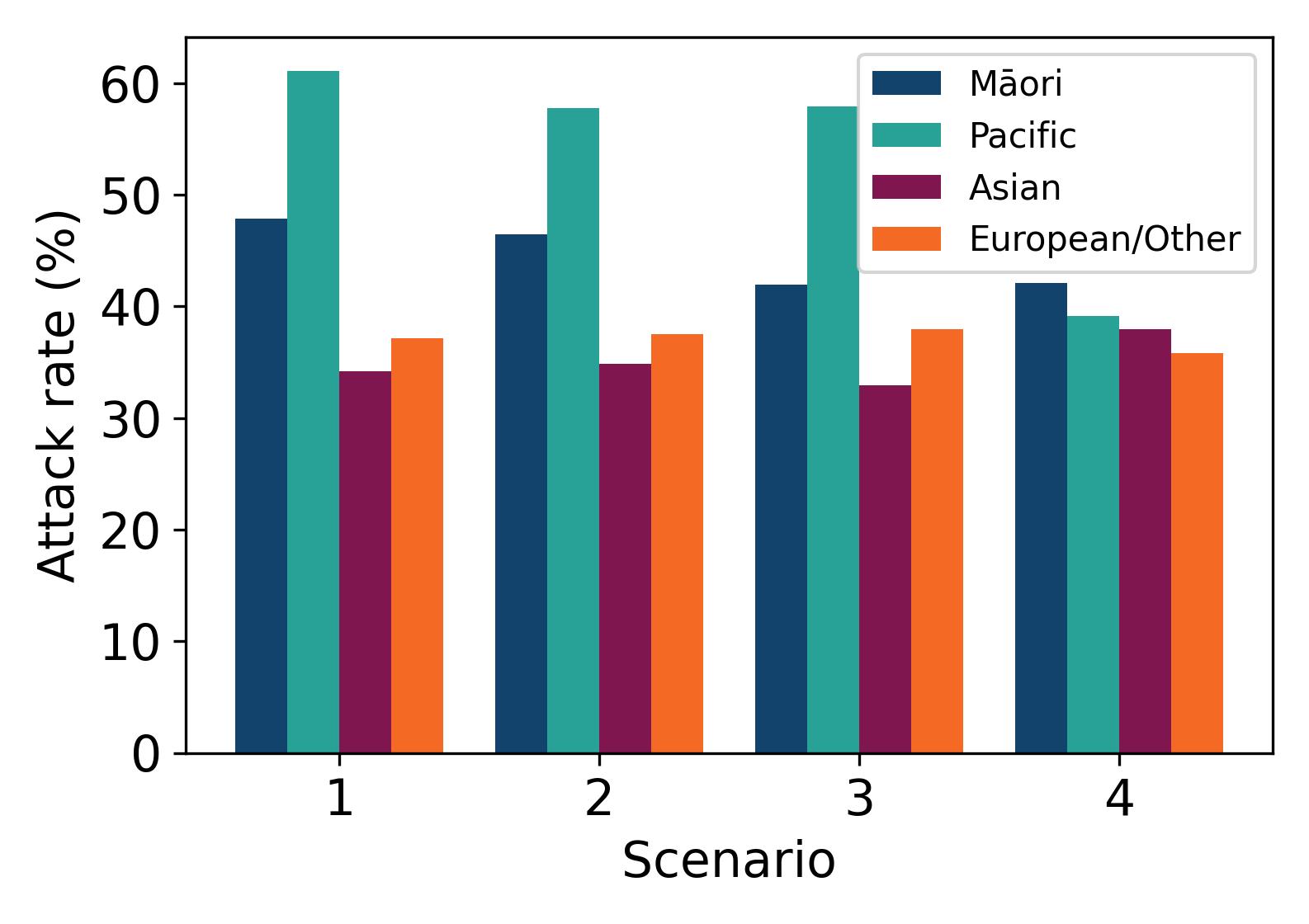}
    \caption{Comparison between the attack rates of the assortative mixing model with 1) ethnicity-specific parameters and fitted value of $\epsilon$, 2) scenario 1 with $\epsilon$ set to zero, 3) scenario 1 with population-averaged vaccination rates, and 4) scenario 1 with population averaged transmission rates (see Table \ref{tab:quantisation}). The assortative model ethnicity-specific parameters and $\epsilon$ were fit to match the attack rates from the prioritised SA2 data with an assumed CAR of 50\%. This meant the attack rates of scenario 1 matched the attack rates found from the data.}
    \label{fig:quantification_barplot_comp}
\end{figure}

Attack rates varied to differing degrees as a result of changes to model parameters representing the scenarios defined in Table \ref{tab:quantisation}. When setting $\epsilon$ to zero (scenario 2) or setting each group's vaccination rate to the population average (scenario 3), there was a relatively small change to the model attack rates (Figure \ref{fig:quantification_barplot_comp}). Conversely, when setting the transmission rates to the population average (scenario 4), there was a significant change to the attack rates; this showed the ethnic disparity in the number of infections was dominated in the model by the disparity in ethnicity-specific transmission rates and not the assortativity value nor the disparity in ethnic vaccination rates. Note that we have only investigated the effect of disparities in vaccination rates on infections. It is likely that disparities in vaccination rates were responsible for larger inequities on mortality and morbidity due to the higher vaccine effectiveness against sever disease and death than against infection or transmission. 

\section{Discussion}

We explored the first Omicron wave during the COVID-19 pandemic in Aotearoa using an ethnicity-stratified SEIR model. We estimated ethnicity-specific transmission rates and an assortativity constant such that the model's attack rates matched those from the COVID-19 case data using three different mixing models: proportionate mixing, assortative mixing, and unconstrained mixing. We then attempted to quantify the relative importance of different drivers in the model.

\subsection{Sensitivity Analysis}

Our model was not sensitive to the spatial resolution of the population data (choice of SA1 or SA2 area level) used to estimate the parameters of the assortative mixing and unconstrained mixing models. Qualitative results did not change when varying the choice of mixing model or assumed value of CAR. This showed that although quantitative estimates of model parameters differ, the qualitative conclusions are robust to the different data sources and assumptions tested. 

\subsection{Mixing model comparison}
All three mixing models fit the cumulative attack rate data exactly because we were using four data points (attack rates in each ethnicity group) to estimate four unknown parameters (transmission rates in each ethnicity group). The assortative mixing model had one additional unknown parameter (the assortativity constant), which was informed by the area-level population data. The mixing models presented make different assumptions about between-group and within-group contact rates. We consider the unconstrained mixing model to be the most representative of real-world contacts, as it is informed directly by empirical data and does not impose any assumptions on the structure of contact matrix. Assortative mixing gave a good approximation for this matrix, however the assumption of a population wide assortativity value was limiting. Proportionate mixing is significantly less representative of real-world mixing. This can be seen in ethnic distributions over Aotearoa being heavily clustered for Pacific and Asian groups (presented in Supplementary Figures 2 and 3).

\subsection{Scenario analysis}

Comparison of different scenarios revealed that, in our model, disparities in vaccination rates between ethnic groups only explain a relatively small part of the observed differences in the ethnicity-specific attack rates, so other factors must have been more important contributors. Likewise, assortative mixing alone is insufficient to explain these differences and can similarly only explain a relatively small difference. Differences between ethnicities in their average transmission rates explained a relatively large part of the observed differences in their attack rates, and therefore transmission rates are likely an important contributor. This is one possible explanation that is consistent with the observed ethnic attack rate disparity.

\subsection{Transmission rate differences}
To reproduce the ethnicity-specific attack rate data from Aotearoa's first Omicron wave with an SEIR model required substantial differences in transmission rates between ethnicities. These transmission rates represent differences in social interactions and not biological responses, i.e., there is not a biological difference between ethnicities. The M\=aori, Pacific, and Asian transmission rates were estimated to be 1.19, 1.87, and 0.85 times higher than European/Other rate under the assumption of proportionate mixing. The approximate magnitude of the relative differences in transmission rates between groups was consistent with findings from Ma et al. \cite{ma_modeling_2021} for New York in 2020; however, the relative differences in our transmission rates were slightly lower than for Ma et al. (2.25, 1.62, 0.86, and 1.28 times the non-Hispanic White rate in New York City for the Hispanic or Latino, non-Hispanic Black, non-Hispanic Asian, and multiracial or other groups, respectively) \cite{ma_modeling_2021}. Like Ma et al., we found that relative differences between transmission rates were smaller under the assumption of assortative mixing as opposed to proportionate mixing. The estimated value of the assortativity constant $\epsilon$ was significantly lower than the Ma et al. value, being 18.3\% instead of 46\% in New York City. Reasons for these differences in the Aotearoa context, covered in the next section, could include workplace, average household occupancy, or age-distribution effects.

\subsection{Possible reasons for differences}
\subsubsection{Jobs and self-isolation}
Work by Chang et al. has shown higher infection rates among disadvantaged racial and socioeconomic groups in part due to their inability to self-isolate as quickly as other groups \cite{chang_mobility_2021}. This was partly due to workplace effects. M\=aori and Pacific Peoples were more likely to work in sectors or roles such as agriculture, transport, and manufacturing jobs \cite{harvey_summary_2023}. While there were no lockdowns during the Omicron wave, these businesses would be operating close to normal levels during the first Omicron wave and would have less opportunity to work from home compared to other sectors. Risk would have also been increased working jobs that are more person-facing. A combination of these factors could have increased the risk and number of interactions M\=aori and Pacific Peoples faced during the first wave of Omicron.

\subsubsection{Deprivation and household crowding}
M\=aori have historically been disadvantaged socio-economically, leading to worse health outcomes \cite{robson_social_2007}. This disparity could be contributing to the disparity in ethnicity-specific transmission rates. In Switzerland, in 2021, adults and young adults living in higher socio-economic positions had fewer contacts than those in lower socio-economic positions. \cite{domenico_individual-based_2025}. This would suggest socio-economic status could partly explain observed differences between ethnicities. M\=aori and Pacific Peoples typically have higher levels of deprivation that lead to an increased risk of hospitalisations from diseases such as influenza \cite{lopez_influenza_2016}. Deprivation is not only associated with the severity of infection but also the incidence rate of diseases. People who live in deprived areas are more likely to live in households with more people in them. Household crowding is a known risk factor for infectious diseases, like COVID-19, and disproportionately affects M\=aori, Pacific, and Asian Peoples \cite{ministry_of_health_analysis_2014,johnson_stocktake_2018}. In 2013, 20\% of M\=aori, 38\% of Pacific Peoples, and 18\% of Asian Peoples lived in crowded households, compared to just 4\% of Europeans. In China, 75\%-80\% of clustered infections were within families, showing high intra-family transmission rates \cite{chen_fangcang_2020}. A meta-analysis of studies has estimated that the risk of household transmission is ten times that of other contacts \cite{lei_household_2020}. Higher household crowding was one factor that probably contributed to higher transmission rates for the M\=aori and Pacific groups. The picture is less clear for the Asian group, as despite the increased rate of household crowding for Asian Peoples, our model estimated lower transmission rates. One possible reason for this was adherence to COVID-19 response measures in which Asian Peoples have been shown to have increased adherence to response measures such as contact tracing and mask wearing, both in Aotearoa \cite{chambers_evaluation_2024,gray_wearing_2020} and overseas \cite{hearne_understanding_2022}.

\subsubsection{Age distribution effects}
M\=aori and Pacific Peoples have a younger age distribution than Asian people and European/Others \cite{bryant_ageing_2003}. As a result, the average age of a case identifying as M\=aori will be substantially lower than for European cases. Some studies have estimated that the majority of household transmissions begin with a child \cite{tseng_smart_2023}, and young adults are frequently estimated to have higher contact rates than older people \cite{prem_projecting_2017,hoang_systematic_2019}. This could contribute towards Māori and Pacific Peoples having higher transmission rates than European people. Susceptibility to COVID-19 in individuals under 20 years of age has been estimated as half that of those over 20, which would mean M\=aori are expected to have fewer infections than European/Others. Another possible issue is that younger cases get reported less, as younger people manifest clinical symptoms less \cite{davies_age-dependent_2020}. This could have led to fewer cases reported in younger populations, which would disproportionately affect M\=aori and Pacific Peoples and imply a lower CAR in these groups than for European/Other people.

\subsubsection{Other possible effects}
In addition, other possible but less documented reasons for differences in transmission rates include differences in frequency of community interaction, method of transportation, and the cultural importance of interacting with extended family, which may be context-specific or hard to quantify. One example of cultural practices that could affect transmission rates was tangihanga, a M\=aori funeral practice, which contains a customary extended ceremony. During the pandemic, these ceremonies had measures to restrict COVID-19 spread, and at some points during the pandemic, tangihanga were not able to be held in any capacity \cite{rangiwai_impacts_2021}. During the Omicron wave, no more than 100 people were allowed into a marae, a traditional M\=aori tribal meeting place, at any one time. Strict COVID-19 measures were used at some marae to prevent transmission, despite this, tangihanga were still potential transmission events during Omicron \cite{cooper_covid_2022}. As cultural activities were altered during COVID-19, it can be difficult to estimate the effect of these activities on disease transmission.

\subsection{Limitations}

\subsubsection{Interaction with other factors}
Our model ignores age-related effects and their interaction with ethnic effects, an obvious avenue for future work, which (as discussed above) could explain some of the observed ethnic disparities in attack rates. The same is true of combining other factors with ethnicity, such as deprivation/socioeconomic status. This is part of a larger limitation that we cannot definitively disentangle differences in transmission rates from differences in reporting rates or other factors. The main difference that drives ethnic disparity in COVID-19 cases is likely to be transmission rates, but there could be other factors that are not included in the model. This is made more difficult to interpret by the fact that these ethnicity-specific transmission rates are proxies for a range of socioeconomic and cultural factors that affect contact patterns and transmission risk. We do not model a causation effect but a correlation between ethnicity and infection rate. These factors can change in complicated ways over time, which means estimates of contact patterns are restricted to the specific context of the first Omicron wave and may not generalise to other situations, e.g. non-pandemic periods or a future pandemic.

\subsubsection{Model fitting and parameters}
In our model, we treated the first Omicron wave in Aotearoa as a SEIR epidemic that ignores differences in time-dependent case rates, which are not well explained by a SEIR model. Our model aimed to be as simple as possible and in this pursuit we used constant parameter values for the ethnicity-specific contact rates. Because our model only had four unknown parameters (transmission rates for the four ethnicity groups), we were able to choose these to provide an exact match with cumulative ethnicity-specific attack rate data. As a consequence, a single model fit does not provide any information about uncertainty; instead we approached this via a sensitivity analysis to investigate how model results were affected by alternative assumptions about mixing patterns, case ascertainment rates, and different choices of area-level population data.

We note that an alternative approach would be to incorporate additional flexibility in the model, via time-varying parameters or the inclusion of other model mechanisms. This would result in a higher-dimensional parameter inference problem, which could be tackled by a Bayesian computational inference method, such as Markov chain Monte Carlo or approximate Bayesian computation, using the full time series of case data as opposed to just the cumulative attack rates. This approach would have the advantage of incorporating parameter uncertainty and enabling systematic comparison of model fit across competing models. However, the drawback of this would be that it would require a much more complex model to obtain a reasonable fit with the time-dependent data, with multiple assumptions about the relevant mechanisms \cite{datta_modelling_2025}. This could risk obscuring the dominant processes affecting observed trends, and suffer from parameter non-identifiability issues that are common in these types of model \cite{roosa_assessing_2019}. Our simplified modelling approach needs to be interpreted with caution due to its limitations. Nevertheless, we believe that it has value as a broad-brush analysis that enables the approximate magnitude of disparities between ethnicity groups and the relative contributions of different effects to be compared in a simplified framework.

Ideally we would have seroprevalence data to complement reported cases to reduce bias present in the data; unfortunately, seroprevalence results were not available for the Omicron wave and the only seroprevalence study in Aotearoa was from earlier in the pandemic \cite{carlton_charting_2021}. We fitted the model to data on reported cases and tested several different assumptions about how these relate to actual infections. We have investigated a scenario with lower CAR for M\=aori and Pacific Peoples, but this still assumes some measure of relationship between each ethnicity's CAR, which may or may not be representative of the real world. This could potentially be partly addressed in future work by including hospitalisations and deaths in the model, which are less sensitive to test-seeking behaviour (though they are sensitive to other differences, e.g. in rates of comorbidity).

\subsubsection{Ethnic groupings}
Our model did not account for mixed ethnicities and used exclusive ethnic groups. The selection of exclusive ethnic groups was a crude approach, as it failed to account for multicultural people who may be affected by different socioeconomic variables and experience different epidemiological trends than any individual ethnic group. Within these groups there is a homogeneity assumption made to simplify modelling that does not account for intra-ethnic variation. One possible way to account for this would be to partition each combination of ethnicities into a separate group. This would represent multicultural individuals better. However, this could lead to group sizes with very few individuals in them that do not give much information, such as individuals who identify with all major ethnic groups.

All considered ethnic groups, excluding M\=aori, are underrepresented in prioritised datasets. For example, when the Ministry of Social Development swapped from a prioritised measure to a total measure, it saw the population identifying as Pacific and European rise from 9\% to 12\% and 39\% to 53\%, respectively \cite{ministry_of_social_development_improving_2021}. Pacific Peoples are also historically underrepresented in the best available population data \cite{sonder_selective_2024}. Both of these can lead to errors in estimates of model parameters, which are difficult to account for.

\subsection{Conclusion}
Our approach described a compartmental ethnicity-stratified model that we used to analyse ethnic transmission rates during the first Omicron wave in Aotearoa. Our model can be further developed to cover some of the limitations outlined and hopefully be used to inform policy on equitable pandemic decisions.

\section*{Acknowledgements}
The authors are grateful to Kevin C. Ma and Stephen Kissler for their useful discussion about contact matrices, to Andrew Sporle, Samik Datta and Nicole Satherley for discussions about modelling infectious disease transmission in different ethnicity groups, and to three anonymous peer reviewers for comments on an earlier version of this manuscript. MJP would like to thank the Banff International Research Station for Mathematical Innovation and Discovery and the organizers and participants of the workshop Mathematical and Statistical Challenges in Post-Pandemic Epidemiology and Public Health (25w5369) held at Casa Matemática Oaxaca in Mexico. 

\section*{Disclosure statement}

We declare that we have no potential conflicts of interest with respect to the investigation, authorship, and publication of this article.

\section*{Funding}

The project ``Improving models for pandemic preparedness and response: modelling differences in infectious disease dynamics and impact by ethnicity'' (TN/P/24/UoC/MP) was funded by Te Niwha, the Infectious Diseases Research Platform – co-hosted by PHF Science and the University of Otago and provisioned by the Ministry of Business, Innovation and Employment, New Zealand. This research was supported by the Marsden Fund grant (24-UOC-020) managed by Royal Society Te Apārangi. VL would like to acknowledge support from the Ng\=ai Tahu Research Centre, University of Canterbury.

\bibliographystyle{tfnlm}
\bibliography{references.bib}

\section{Appendix - supplementary}
\subsection{Statistical area information}

Table \ref{table:SA} contains various summary statistics about the statistical area datasets used in parameter fitting.

\begin{table}[ht]
\centering
\begin{tabularx}{0.93\textwidth}{c c c c c c c} 
 \toprule
\multirow{2}{*}{} & \multirow{2}{*}{Total} & \multirow{2}{*}{Mean} & \multirow{2}{*}{Median} & \multirow{2}{*}{SD} & \multirow{2}{*}{Range} & Interquartile \\ &&&&&&range\\
\midrule
\multicolumn{7}{l}{\textbf{SA1 Total measure}}\\
\midrule
M\=aori & 775410 & 26.5 & 18.0 & 28.2 & 0 - 435 & 9 - 33\\
Pacific Peoples & 381576 & 13.0 & 6.0 & 25.4 & 0 - 315 & 0 - 12\\
Asians & 707499 & 24.2 & 12.0 & 33.4 & 0 - 789 & 3 - 33\\
Europeans/Others & 3424572 & 117.0 & 114.0 & 49.8 & 0 - 825 & 84 - 147\\
All ethnicities & 5289057 & 180.7 & 174.0 & 60.4 & 27 - 1188 & 141 - 213\\
\midrule
\multicolumn{7}{l}{\textbf{SA2 Total measure}}\\
\midrule
M\=aori & 775875 & 362.6 & 252.0 & 355.7 & 0 - 3477 & 138 - 472\\
Pacific Peoples & 381711 & 178.4 & 60.0 & 381.4 & 0 - 3501 & 24 - 138\\
Asians & 707541 & 330.6 & 126.0 & 467.6 & 0 - 4530 & 45 - 414\\
Europeans/Others & 3425964 & 1600.9 & 1539.0 & 827.3 & 9 - 4230 & 995 - 2136\\
All ethnicities & 5291091 & 2472.5 & 2395.5 & 1243.7 & 33 - 6099 & 1518 - 3319\\
\midrule
 \multicolumn{7}{l}{\textbf{SA2 Priority measure}}\\
\midrule
M\=aori & 888840 & 412.1 & 290.0 & 396.6 & 0 - 3780 & 160 - 540\\
Pacific Peoples & 359480 & 166.7 & 50.0 & 376.7 & 0 - 3440 & 20 - 120\\
Asians & 820580 & 380.4 & 150.0 & 534.0 & 0 - 5760 & 50 - 470\\
Europeans/Others & 3048470 & 1413.3 & 1330.0 & 821.0 & 10 - 7600 & 830 - 1910\\
All ethnicities & 5117370 & 2372.4 & 2300.0 & 1215.5 & 10 - 10120 & 1470 - 3170\\
\bottomrule
\end{tabularx}
\caption{Table of various statistical measures of the ethnic populations within the SA datasets, where SD is the standard deviation of the populations, and interquartile range is the range between the 75th and 25th quartile of the population sizes.}
\label{table:SA}
\end{table}

\subsection{Transmission rate values}

Values of the transmission rates fit to the SA1 and SA2 total measure population projections can be seen in Table \ref{table:parameter_values_assortative}.

\begin{table}[ht] 
\centering
{\begin{tabular}{lc c c c c} 
\toprule
 &  & $a_k$ & & &\\ \cmidrule{3-6}
\textbf{CAR} & $\epsilon$ & M\=aori & Pacific & Asian & European/Other\\
\midrule
SA1\\
\midrule
40/60 & 0.162 & 0.617 & 1.019 & 0.281 & 0.328\\
40 & 0.188 & 0.556 & 0.916 & 0.403 & 0.476\\
50 & 0.195 & 0.496 & 0.742 & 0.380 & 0.444\\
60 & 0.198 & 0.465 & 0.668 & 0.366 & 0.425\\
\midrule
SA2\\
\midrule
40/60 & 0.122 & 0.622 & 1.039 & 0.275 & 0.321\\
40 & 0.155 & 0.558 & 0.928 & 0.400 & 0.473\\
50 & 0.160 & 0.498 & 0.753 & 0.377 & 0.441\\
60 & 0.163 & 0.467 & 0.678 & 0.363 & 0.423\\
\bottomrule
\end{tabular}}
\caption{Table of transmission rate values fit using the assortative mixing model for total population data.}
\label{table:parameter_values_assortative}
\end{table}

\subsection{Column summation property}
The columns of our matrix must sum to the transmission rates (described by the below equation).

\begin{align*}
    \sum_i C_{ij} = a_j
\end{align*}

We refer to this as the column summation condition. This condition requires the sum of contacts an individual has with each ethnic group equals the contact rate of the individual's ethnicity.

\subsubsection{Frequency-based assumption}

The frequency-based empirically estimated contact matrix is defined via the following equation,

\begin{align}
    T_{ij}' & = \sum_l\frac{a_ia_jN_{i,l}N_{j,l}}{\sum_ka_kN_{k,l}} 
\end{align}

This leads to the following definition of the transmission matrix

\begin{align*}
    C_{ij,\,\text{frequency}} & = \frac{1}{N_j}\sum_l\frac{a_ia_jN_{i,l}N_{j,l}}{\sum_k a_k N_{k,l}}
\end{align*}

where the index $l$ represents a block and $N_{kl}$ represents an ethnicity's population within that block.

This matrix does conserve the meaning of the transmission rates as seen by the following derivation

\begin{align*}
    \sum_i C_{ij,\,\text{frequency}} & = \sum_i\frac{1}{N_j}\sum_l\frac{a_ia_jN_{i,l}N_{j,l}}{\sum_k a_k N_{k,l}}\\
    & =\frac{1}{N_j}\sum_l\frac{a_jN_{j,l}\sum_ia_iN_{i,l}}{\sum_k a_k N_{k,l}}\\
    & = \frac{1}{N_j}\sum_l a_jN_{j,l}\\
    & = a_j
\end{align*}

\subsubsection{Density-based assumption}

The density-based empirically estimated contact matrix is defined by Ma et al. via the following equation,

\begin{align*}
    \beta_{ij,\,\text{density}} & = \frac{\sum_k a_k N_k}{\sum_{i,j,l}a_ia_jN_{i,l}N_{j,l}}\sum_l a_ia_jN_{i,l}N_{j,l}
\end{align*}

This matrix does not satisfy the column summation condition.

\subsection{Parameter Variation}
\begin{figure}[ht]
    \centering
    \includegraphics[width=0.8\linewidth]{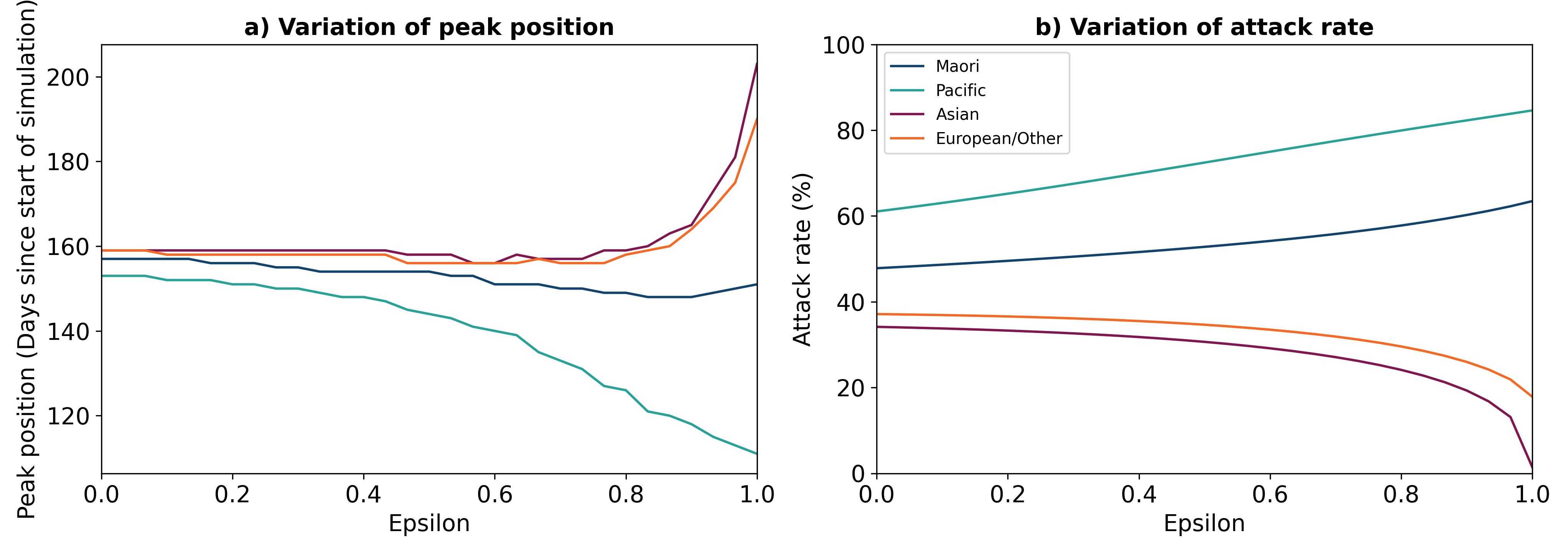}
     \caption{a) Plot of the time of the peak position of new infections for each ethnicity while varying $\epsilon$. The transmission vector, $\boldsymbol{a}$, was refit to the attack rates from the data for every value of epsilon plotted, such that every point on the graph was from a model that matched the attack rates from the data. b) Plot of attack rate values with variation of $\epsilon$. The transmission vector, $\boldsymbol{a}$, was fit to the attack rates from the data with an assumed CAR of 50\% and no assortativity (i.e., pure proportional mixing).}
    \label{fig:epsilon_variation}
\end{figure}

The transmission rates were held constant to the values fit to the proportionate mixing scheme while $\epsilon$ varied. As can be seen in Figure \ref{fig:epsilon_variation}, the disparity in attack rate between groups widened as $\epsilon$ increased. This was expected, as when epsilon approached one, different ethnicities' equations decoupled into separate SEIR models, which had no dependence on other ethnicities transmission rates. $\epsilon$ was also varied while the transmission rates were refit to the attack rate data (under the assumed value for CAR). It would not be qualitatively reasonable if the gap between the times of the peak positions was too large and values over 0.6 tend to not have qualitatively reasonable gaps in peak timing.

\subsection{Additional plots}

\begin{figure}
\centering
\subfloat{%
\resizebox*{0.44\textwidth}{!}{\includegraphics{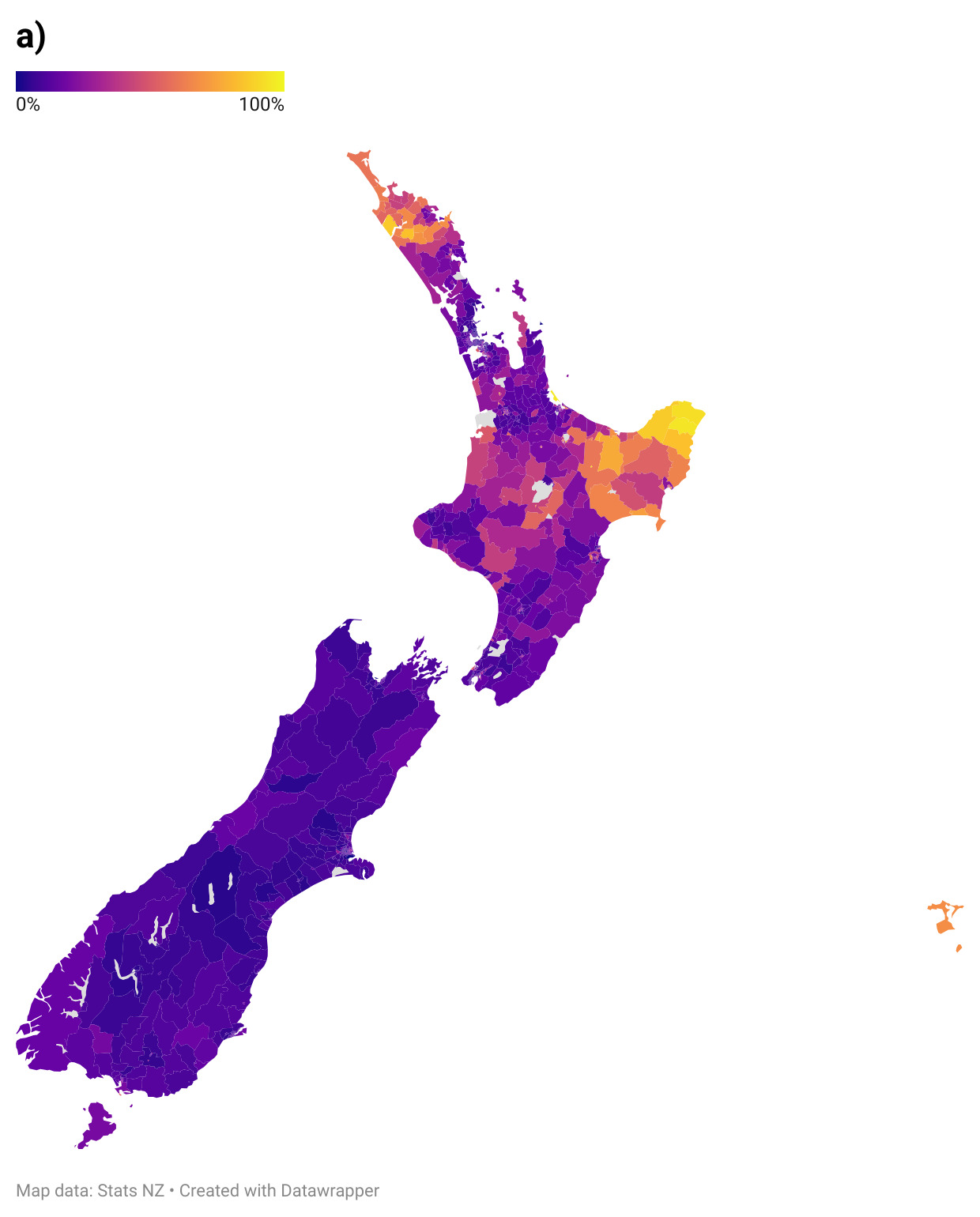}}}\hfill
\subfloat{%
\resizebox*{0.44\textwidth}{!}{\includegraphics{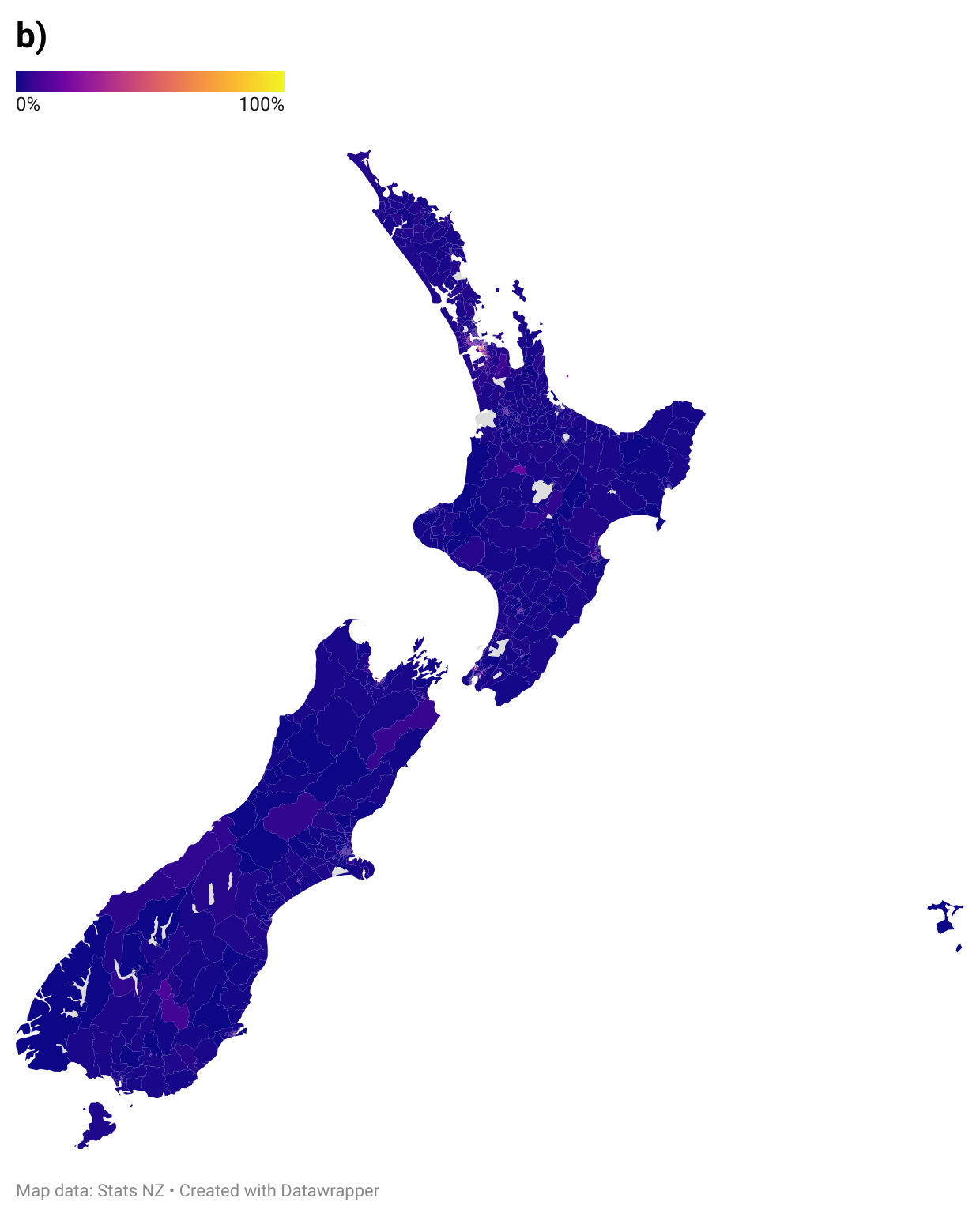}}}\\
\subfloat{%
\resizebox*{0.44\textwidth}{!}{\includegraphics{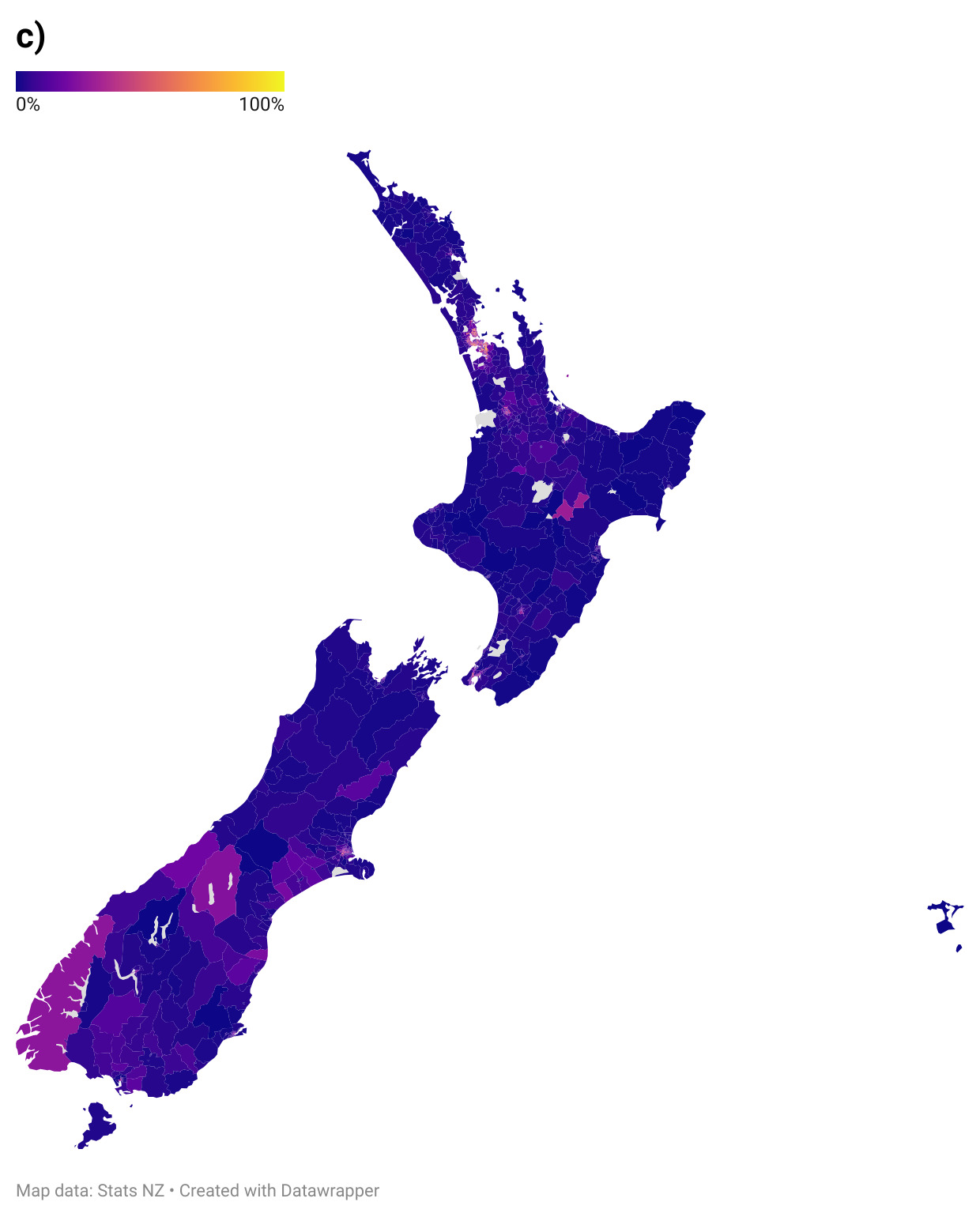}}}\hfill
\subfloat{%
\resizebox*{0.44\textwidth}{!}{\includegraphics{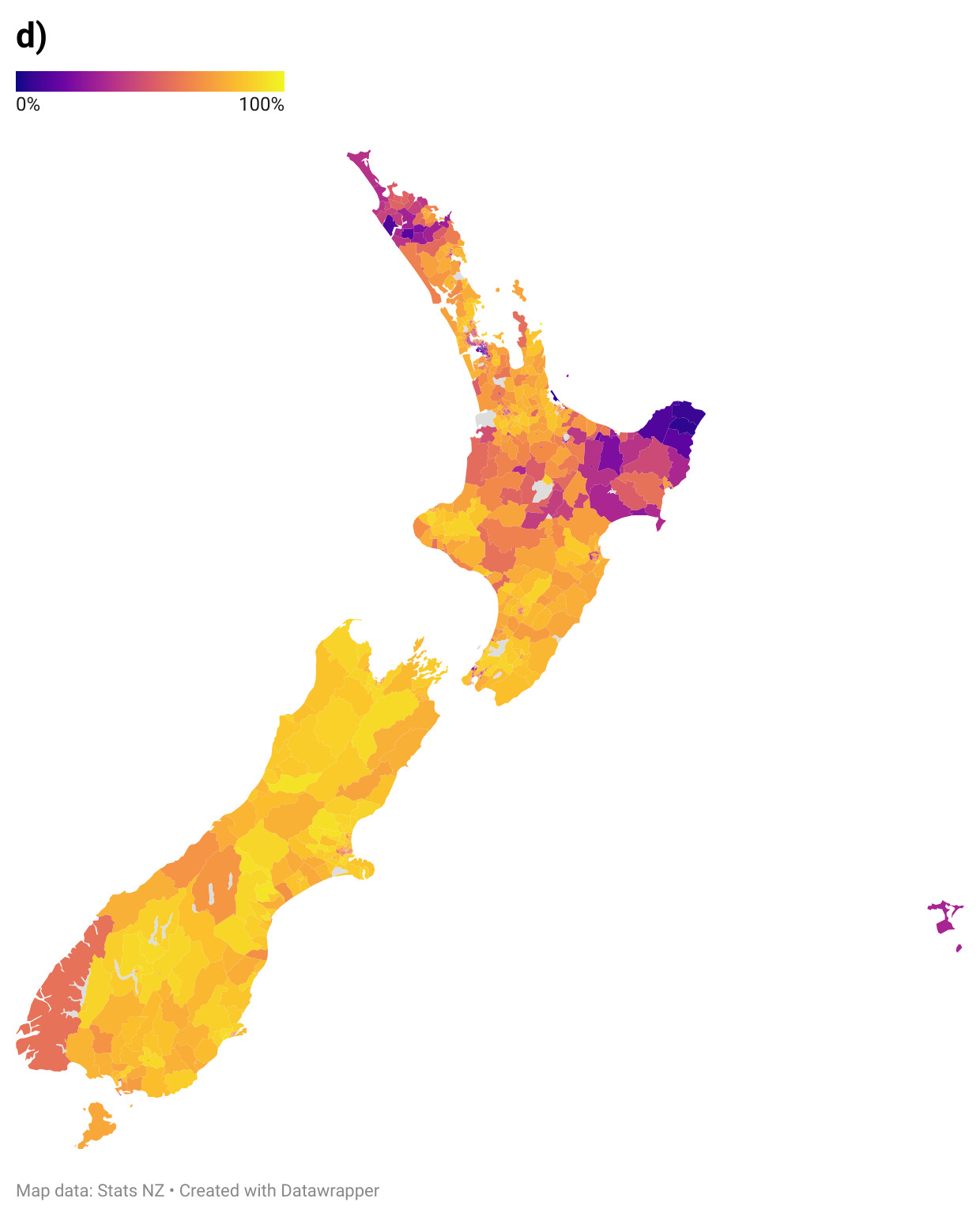}}}
\caption{Proportion of each SA2 that is a) M\=aori, b) Pacific, c) Asian, and d) European/Other.} \label{fig:SA2_maps}
\end{figure}

\begin{figure}
\centering
\subfloat{%
\resizebox*{0.44\textwidth}{!}{\includegraphics{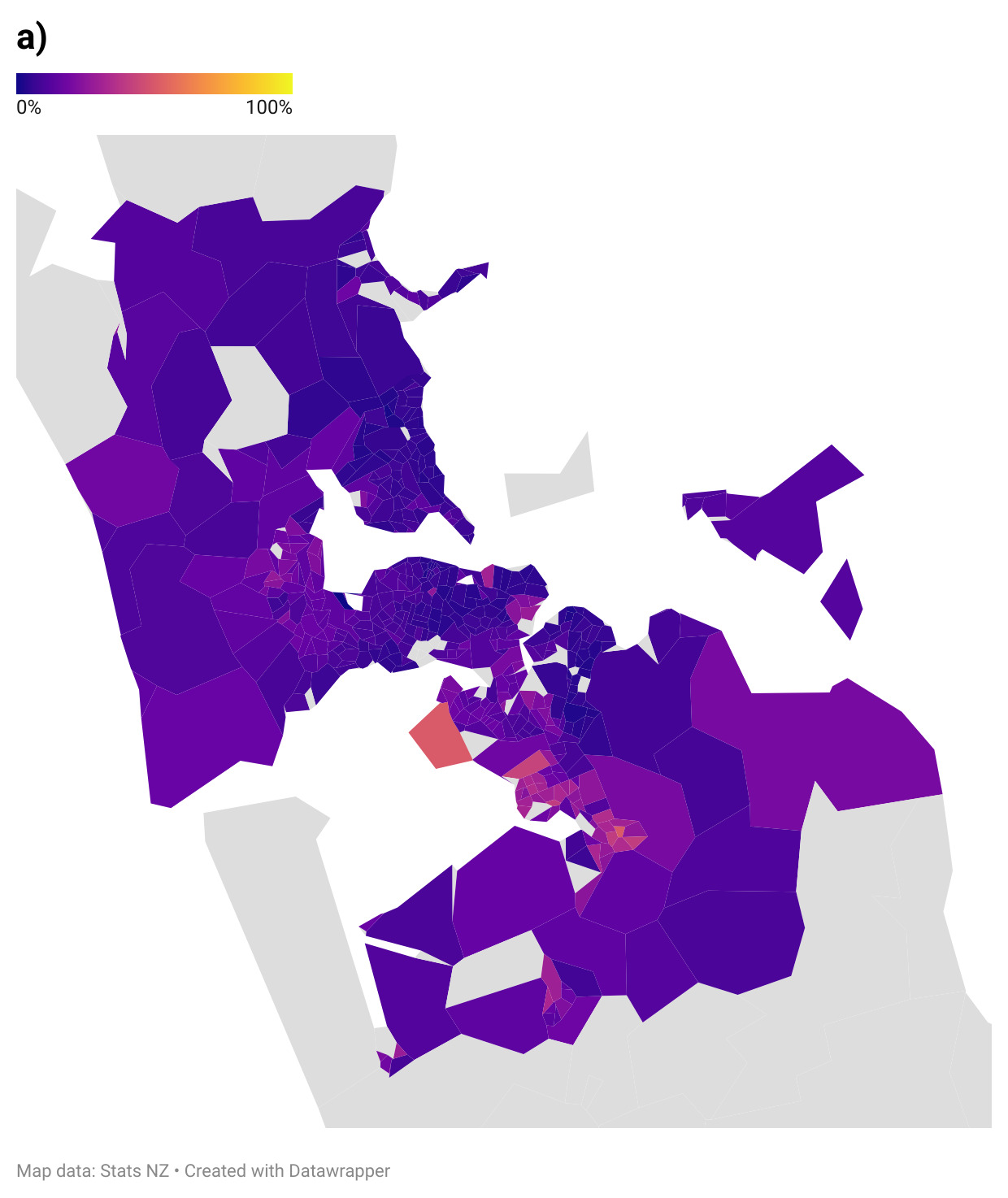}}}\hfill
\subfloat{%
\resizebox*{0.44\textwidth}{!}{\includegraphics{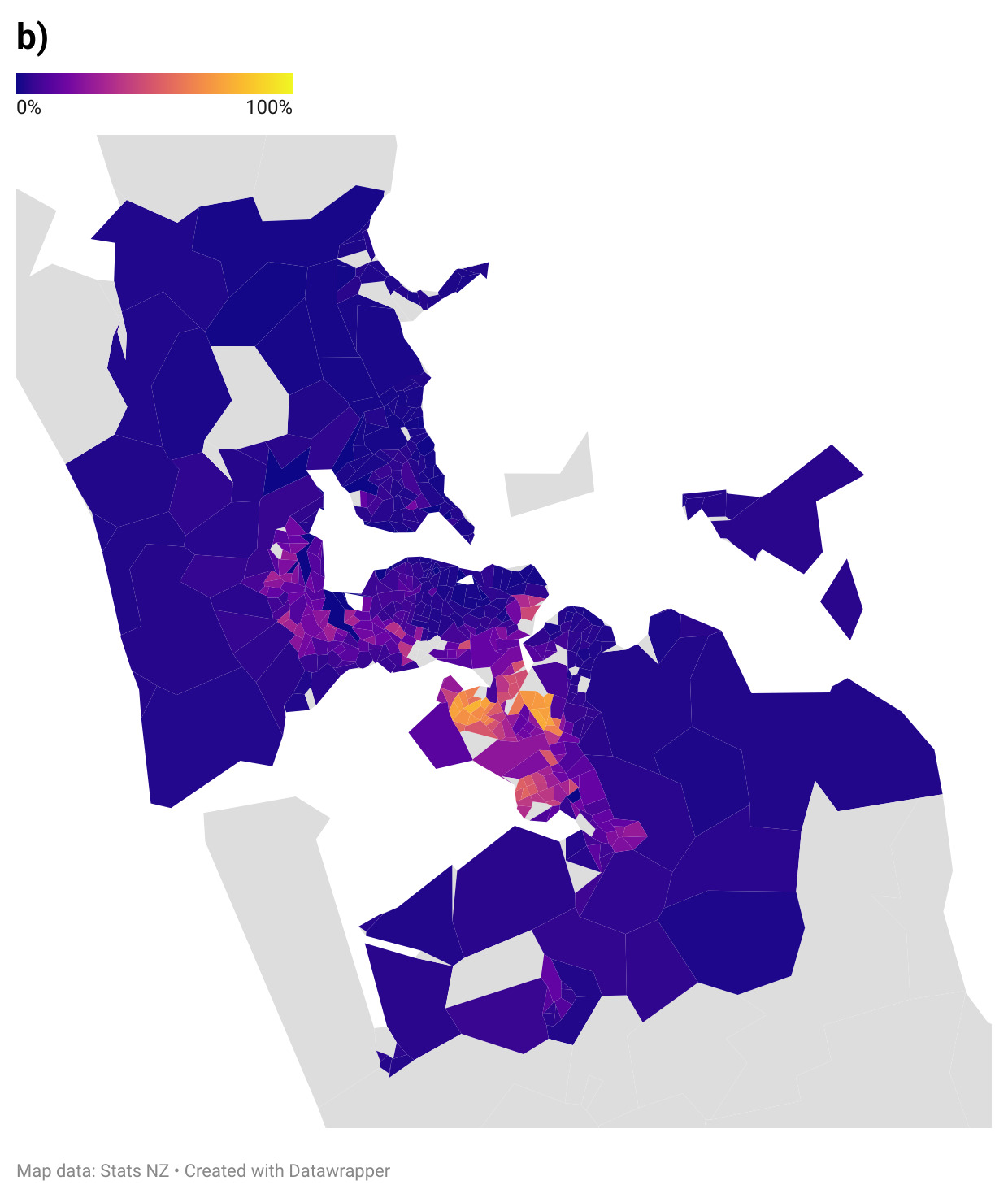}}}\\
\subfloat{%
\resizebox*{0.44\textwidth}{!}{\includegraphics{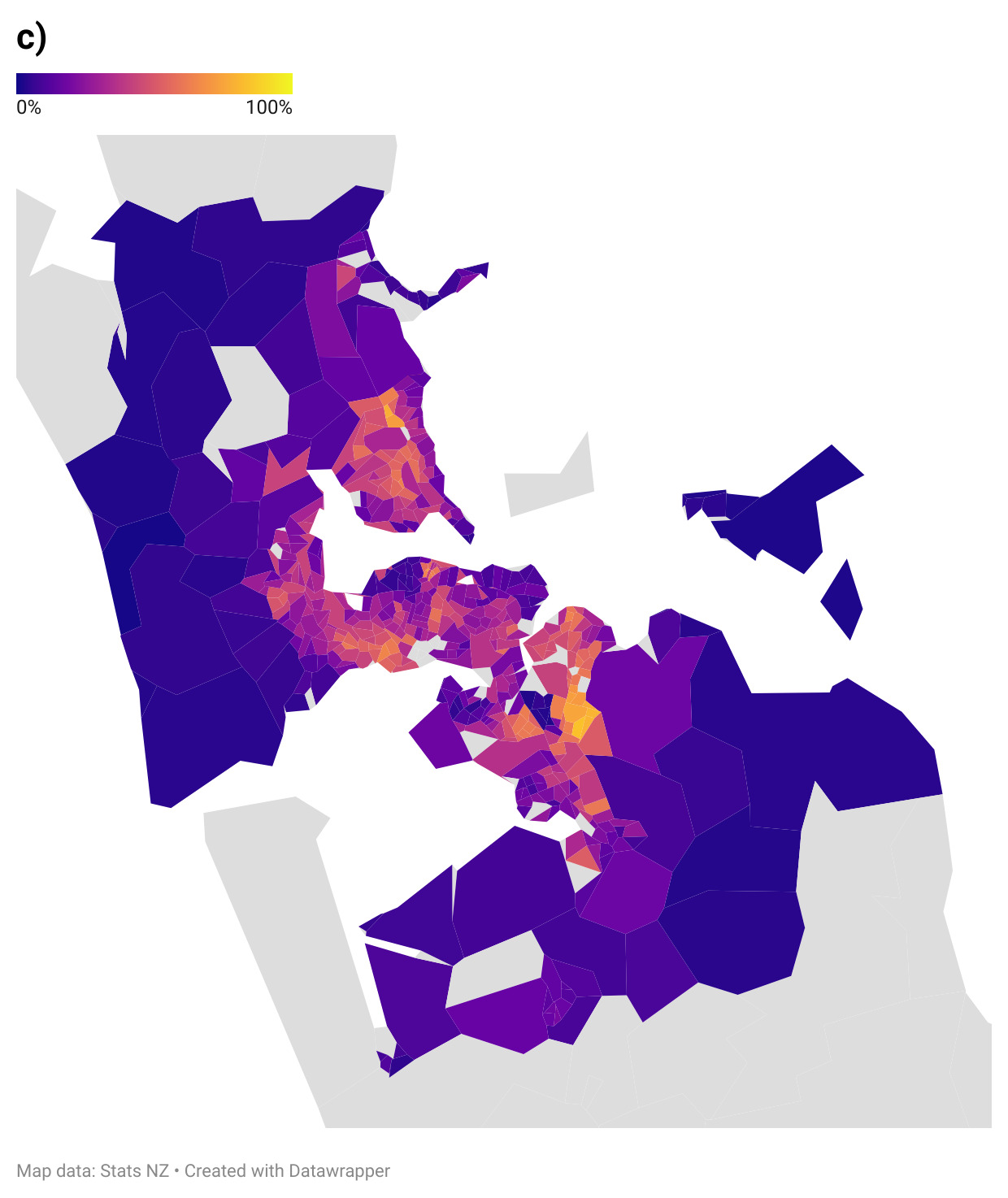}}}\hfill
\subfloat{%
\resizebox*{0.44\textwidth}{!}{\includegraphics{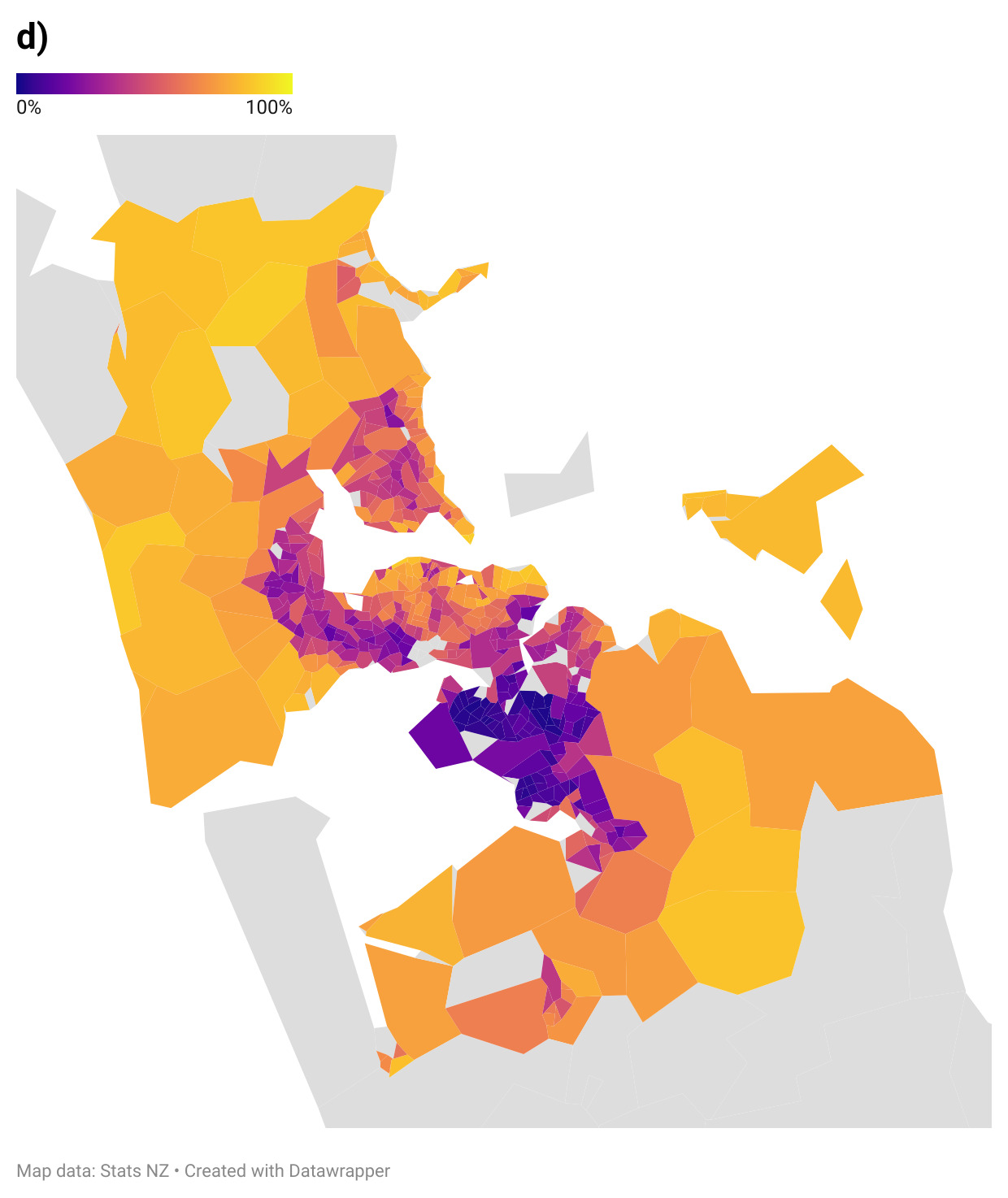}}}
\caption{Proportion of each SA2 that is a) M\=aori, b) Pacific, c) Asian, and d) European/Other in the majority of Auckland SA2 regions.} \label{fig:SA2_auckland_maps}
\end{figure}

Maps of the ethnic composition of each SA2 region in Aotearoa can be seen in Figure \ref{fig:SA2_maps}. There is high ethnic clustering of the Pacific and Asian groups in cities, particularly Auckland (Figure \ref{fig:SA2_auckland_maps}.

\begin{figure}[ht]
    \centering
    \includegraphics[width=0.9\linewidth]{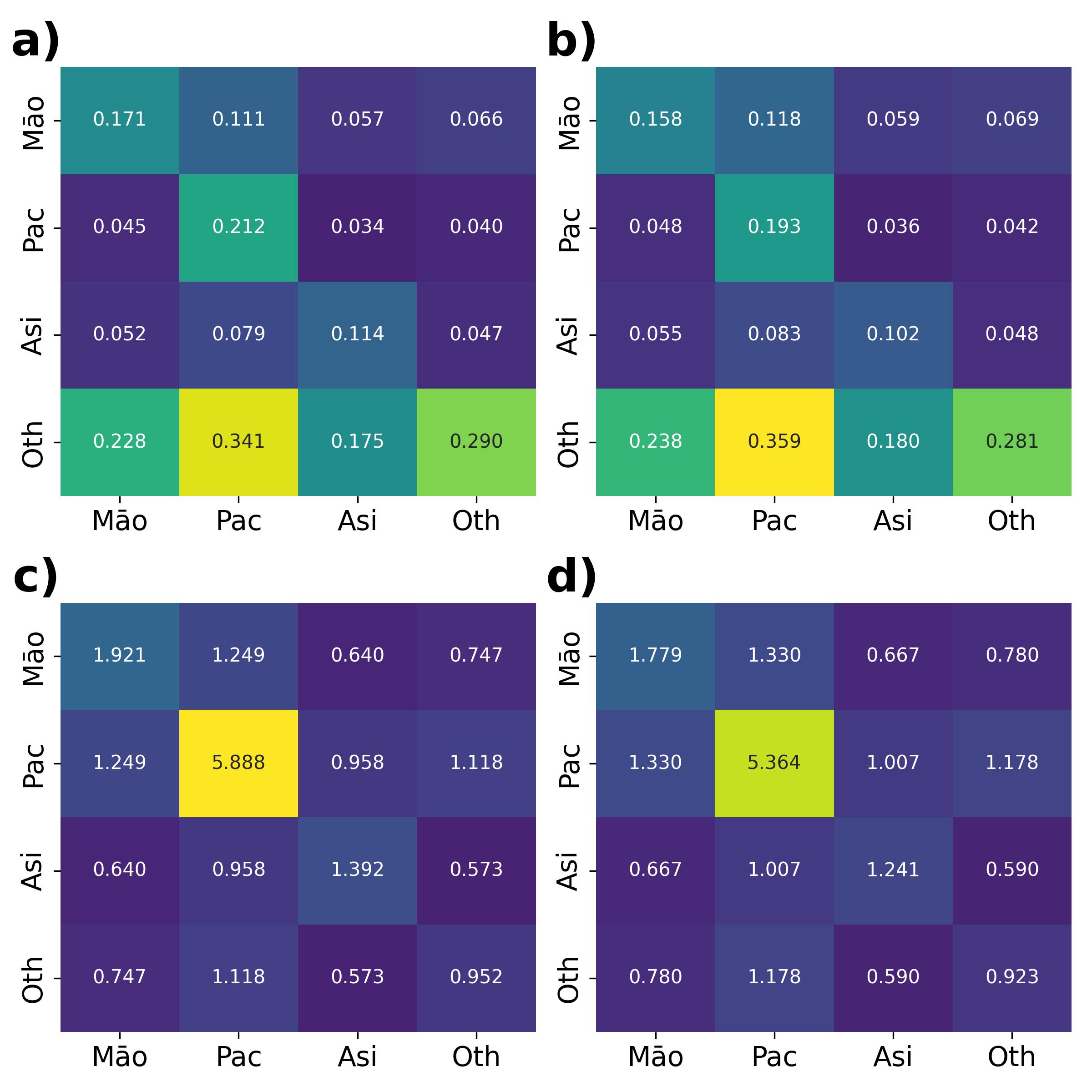}
    \caption{Comparison between the assortative mixing transmission matrices fit using a) SA1 total population data and b) SA2 total population data. Matrices were fit to the attack rate data with an assumed CAR of 50\%.}
    \label{fig:SA_heatplot}
\end{figure}

\begin{figure}
    \centering
    \includegraphics[width=0.98\linewidth]{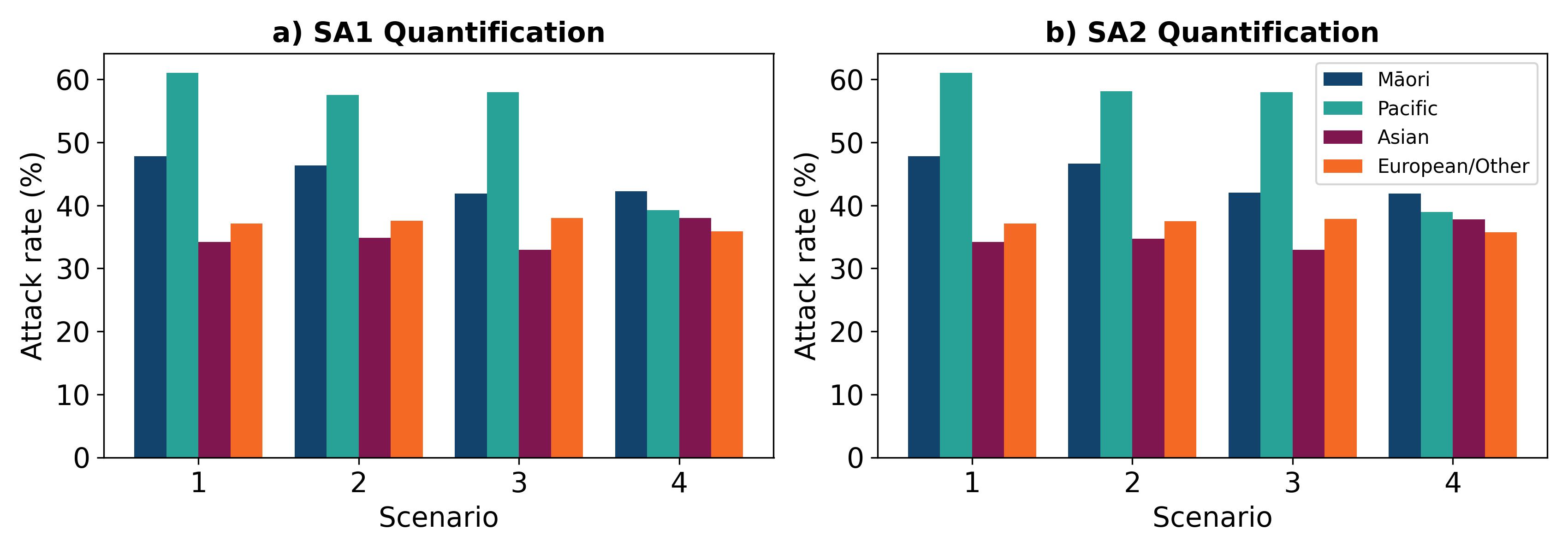}
    \caption{Quantification analysis using a) the SA1 and b) the SA2 total measure population data. The four scenarios considered are:  1) ethnicity-specific parameters and assortativity, 2) no assortativity, 3) population-averaged vaccination rates, 4) population-averaged transmission rates.}
    \label{fig:SA_quantification}
\end{figure}

Figures \ref{fig:SA_heatplot} compare the assortative transmission matrices fit using the total measure population data. Figure \ref{fig:SA_quantification} shows quantification analysis repeated for total measure statistical area population data.

\end{document}